\documentclass[%
 twocolumn,
 reprint,
 superscriptaddress,
 nofootinbib,
 amsmath,
 aps, pra,
]{revtex4-2}

\usepackage{color}
\usepackage{graphicx}
\usepackage{hyperref}

\DeclareMathOperator{\Tr}{\text{Tr}}

\begin{document}

\title{Bit-flip errors in dissipative cat qubits: Second-order perturbation theory}

\author{Kirill S. Dubovitskii}
\affiliation{Univ.~Grenoble Alpes, CNRS, LPMMC, 38000 Grenoble, France}

\begin{abstract}
Dissipative cat qubits are known for the exponential suppression of the bit-flip rate. However, there is significant discrepancy between experimental measurements and analytical predictions of the strength of the bit-flip suppression. In this paper we resolve this discrepancy for some of the perturbations, by developing a second-order perturbation theory on top of a nonlinear dissipative Lindbladian. Following this scheme, we derive an analytical expression for the exponentially small bit-flip rate due to single-photon loss, which shows good agreement with numerical simulations. We also apply our scheme to other perturbations, such as frequency detuning and the Z gate, and find the corresponding bit-flip rates, which also show good agreement with the numerical simulation.

\end{abstract}

\maketitle
\section{Introduction}
Building a fault-tolerant quantum computer is extremely challenging. Over recent decades, various qubit architectures have been proposed, including plenty of superconducting qubits~\cite{Kjaergaard2020}. Currently, qubits with intrinsic protection against one type of error are considered promising~\cite{Gyenis2021}. Such qubits significantly reduce the overhead of error correction protocols~\cite{Aliferis2008,Tuckett2018,Tuckett2019,PhysRevLett.124.130501}. An example of such a qubit is the Schrödinger cat qubit~\cite{Mirrahimi2014}. It is encoded in a bosonic mode such that the computational states are separated in the phase space. The separation parameter is the average photon number $\alpha^2 \gtrsim 1$ of the mode, known as the cat-state size. The bit-flip error rate of a cat qubit is exponentially suppressed with increasing cat-state size, while phase flips scale linearly, making bit flips much less probable than phase flips.

To meet quantum error correction thresholds, both bit-flip ($X$ error) and phase-flip ($Z$ error) rates must be sufficiently low~\cite{Guillaud2019}. While analytical expressions for phase-flip errors have been found for most noise channels and gates and align well with experiments~\cite{Guillaud2021, Chamberland2022}, the understanding of bit-flip errors remains incomplete.
In particular, the bit-flip rate of the cat qubit due to photon loss, a dominant noise channel~\cite{Leghtas2015, Lescanne2020, Reglade2024, Berdou2023}, is not fully explained. The available analytical expression~\cite{Gautier2022} predicts an exponential bit-flip rate proportional to $ e^{-4\alpha^2}$, arising from the perturbation matrix element between the computational states, which corresponds to direct probability transfer inside the computational subspace. However, observed suppression is significantly weaker~\cite{Leghtas2015, Lescanne2020, Berdou2023, Reglade2024}. Numerical studies~\cite{Guillaud2023, Regent2023} suggest bit-flip suppression proportional to $e^{-c\alpha^2}$ with $c\gtrsim 2$, contradicting the analytical prediction~\cite{Gautier2022}.
The situation with other error sources is also unclear. For example, there are no analytical expressions for the bit-flip probability due to gate implementation and frequency detuning~\cite{Lescanne2020}.

The complexity behind the bit-flip analysis lies in the fact that bit-flip rates are exponentially suppressed and thus nonperturbative in $1/\alpha^2$, while the phase-flip rate can be found perturbatively in the inverse cat-state size. Another technical difficulty is that the dissipative cat qubit is unconventional as its unperturbed dynamics is governed by a Lindbladian $\mathcal{L}_0$ rather than a Hamiltonian. This dissipative dynamics stabilizes the computational states, attracting any probability that leaks out of the computational subspace due to spurious noises. The key question for bit flips is how this leaked probability is distributed among the computational states during stabilization.
Assuming noises or gates are weak compared to the leading dynamics, error rates are described as perturbations $\mathcal{L}_1$ of the zeroth-order Lindbladian $\mathcal{L}_0$. Available analytical expressions for bit-flip rates are first-order perturbative results, but, as we said, for some common perturbations these expressions fail to explain the bit-flip rates observed in experiments. To understand why, we need to examine the probability transfer between computational states caused by perturbations.

Besides the direct probability transfer, there is another mechanism specific to dissipative cat qubits.
It
involves the probability that leaks to noncomputational
states and is subsequently returned by the dissipative
dynamics. The probability to leave one computational state and return to the other is proportional to
$e^{-2\alpha^2}$~\cite{Mirrahimi2014, Thompson2022} (see Appendix~\ref{app:leakingperturbations}).
Due to the weaker dependence on $\alpha^2$, the second mechanism dominates for larger cat-state sizes. However, to capture this mechanism, one has to develop the perturbation theory to the order in which the perturbation starts to cause leakage. 

Thus, for some perturbations like dephasing (frequency fluctuations) or photon gain, the first-order bit-flip expressions are complete~\cite{Mirrahimi2014,Dubovitskii2024}. Indeed, their jump operators $\hat{a}^\dagger \hat{a}$ and $\hat{a}^\dagger$ acting on a computational state wave function produce amplitudes in the noncomputational space. For some other perturbations, which do not cause leakage in the first order like photon loss or Hamiltonian type of the form $-i[\hat{V}, \hat{\rho}]$, the first-order calculation does not include the noncomputational transfer mechanism; therefore, one has to study the bit-flip rate in the second order to recover the leading exponential. However, to study the second order, one generally needs to be able to invert the unperturbed Lindbladian $\mathcal{L}_0$, which remains an unresolved task for the cat qubits.

\begin{figure}
    \centering
    \includegraphics[scale=0.5]{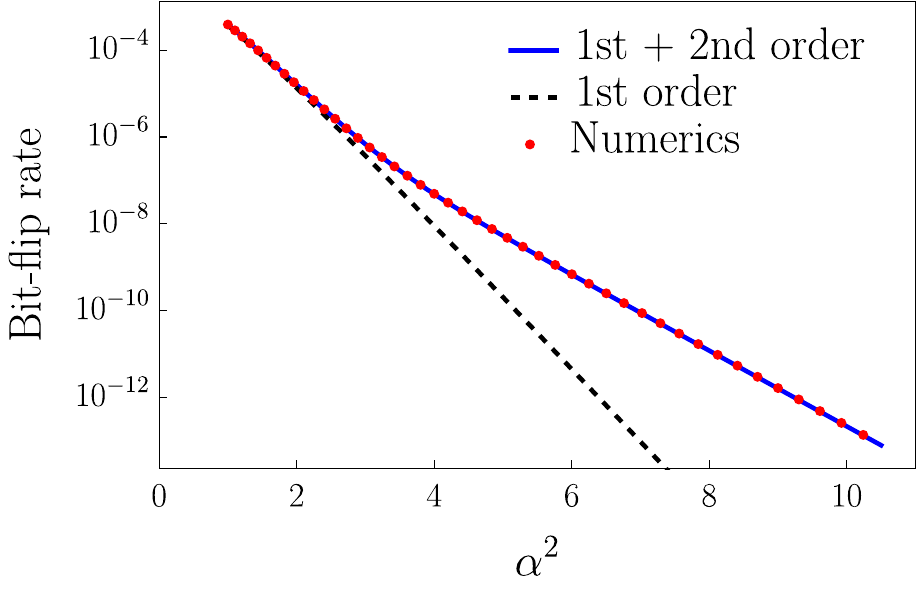}
    \caption{Bit-flip rate in units of $\kappa_2$ due to the single photon loss of strength $\kappa_1 = \kappa_2/100$. The crossover between two exponentials happens at $\alpha^2\approx \ln{(\kappa_2/\kappa_1)}/2$. Remarkably, after accounting for the second order, the bit-flip-rate expression agrees well with the numerical simulation. }
    \label{fig:bitflip}
\end{figure}

In this paper we circumvent this problem by evaluating the action of $\mathcal{L}_0^{-1}$ on the subset of virtual states involved in the second-order contribution to the bit-flip rates due to  perturbations, which do not cause leakage in the first order, but cause it in the second order. In particular, we derive the bit-flip rate of the cat qubit due to photon loss, as well as for the Z gate and detuning. To support our analytical results, we perform numerical simulations that show good agreement with the analytics across the entire range of cat-state size $\alpha^2$ (see Fig.~\ref{fig:bitflip}).
There are perturbations that do not cause leakage in the first order nor in the second order. We show in Appendix~\ref{app:completelyNonLeaking}, that such perturbations do not cause leakage at all. They are trivial in the sense that one finds the exact bit-flip rate by diagonalizing their action truncated to the computational subspace.

\section{Model and summary of the main results}
In order to present our quantitative results we now introduce the model. Briefly, the main idea of the dissipative cat qubit is to create two-photon dissipation of the memory mode (in which the cat qubit is encoded). Usually, such dissipation is achieved by engineering a two-photon exchange between the memory mode and a lossy buffer mode~\cite{Lescanne2020,Berdou2023,Reglade2024, Marquet2024}. Then application of a resonant drive to the buffer mode stabilizes the computational subspace in the memory mode. After adiabatic elimination of the lossy mode degree of freedom, the dynamics of the memory mode is governed by the master equation in the rotating frame, 
\begin{equation} \label{eq:master_eq}
    \frac{d\hat{\rho}}{dt}=\mathcal{L}_0\hat{\rho}\equiv\kappa_2\mathcal{D}\left[\hat{a}^{2}-\alpha^{2}\right]\hat{\rho},
\end{equation}
where $\hat{a}$ is a bosonic ladder operator in the Hilbert space of a harmonic oscillator, $\hat{\rho}$ is the density matrix of the memory mode existing in the same space, $\alpha$ is a number, that can be taken positive without loss of generality, $\kappa_{2}$ is a two-photon dissipation rate, and finally, $\mathcal{D}[\hat{L}]\hat{\rho}\equiv\hat{L}\hat{\rho}\hat{L}^{\dagger}-\frac{1}{2}\hat{L}^{\dagger}\hat{L}\hat{\rho}-\frac{1}{2}\hat{\rho}\hat{L}^{\dagger}\hat{L}$. The Lindblad superoperator $\mathcal{L}_0$ has four stationary eigenoperators $\vert{\sigma\alpha}\rangle\langle {\sigma'\alpha}\vert$ with $\sigma, \sigma'=\pm1$, where $\vert{\pm\alpha }\rangle$ are the coherent states of the harmonic oscillator. This stable computational subspace can be seen as the qubit Bloch sphere. If spontaneous leakage occurs, the probability outside the computational subspace flows back to it in a way described by the master equation~\eqref{eq:master_eq}. 

Let the $x$ axis of the Bloch sphere lie along the so-called Schrödinger cat states,
\begin{equation}\label{eq:catstates}
\vert\mathcal{C}_{\alpha}^{\pm}\rangle=\frac{\vert\alpha\rangle\pm\vert{-\alpha}\rangle}{\sqrt{2\left(1\pm e^{-2\alpha^{2}}\right)}}.
\end{equation}
For the computational states (the $z$ axis), we pick
\begin{align}
    \vert 0_c \rangle = \frac{\vert\mathcal{C}_{\alpha}^{+}\rangle + \vert\mathcal{C}_{\alpha}^{-}\rangle}{\sqrt{2}},\phantom{s}
    \vert 1_c \rangle = \frac{\vert\mathcal{C}_{\alpha}^{+}\rangle - \vert\mathcal{C}_{\alpha}^{-}\rangle}{\sqrt{2}}.
\end{align}
For large $|\alpha|^2 \gg 1$ the states $\vert 0_c\rangle\approx \vert \alpha\rangle$ and $\vert{1_c}\rangle\approx \vert {-\alpha}\rangle$ are well-separated in phase space, suppressing the matrix elements of any local noise between them.

Undesired processes manifest as perturbations to the dynamics \eqref{eq:master_eq}. The computational subspace is no longer invariant and Bloch vector components decay slowly. The decay rate of the $z$ component constitutes the bit-flip error, while the decay rates of the $x$ and $y$ components describe the phase-flip errors. These decay rates are given by the real part of the perturbed eigenvalues $\lambda_x$, $\lambda_y$, and $\lambda_z$ of the corresponding Bloch sphere component. Later in the paper we will calculate the bit-flip rate, formally defined as $\Gamma_{0\leftrightarrow1}=-\text{Re }\lambda_z /2$. Then the first- or second-order corrections to the bit-flip rate are given by the corresponding perturbative corrections to the eigenvalue $\lambda_z$. As we mentioned before, the dominant error channel in most realizations of the cat qubits is the photon loss, which is accounted for by adding the term $\mathcal{L}_{1}=\kappa_1\mathcal{D}\left[\hat{a}\right]$ to the master equation~\eqref{eq:master_eq}. The photon loss does not cause leakage in the first order due to $\hat{a}\vert {\pm \alpha} \rangle = \pm \alpha \vert {\pm \alpha} \rangle$. To formalize it, we quantify leakage $l(t)$ as it is done in Ref.~\cite{Regent2023},
\begin{equation}
    l(t) = 1 -  \sum_\sigma \langle \mathcal{C}_\alpha^\sigma\vert \hat{\rho}(t) \vert \mathcal{C}_\alpha^\sigma\rangle\equiv 1-\Tr_\|\left[\hat{\rho}(t)\right],
\end{equation} 
where we introduced the trace over the computational space, denoted by $\Tr_\|$. If $l(t) = 0$, then there is no population outside the computational space.  
Now we use this definition to characterize whether a perturbation causes leakage in the first order. Let us initialize the qubit in the computational state $\hat{\rho}(0)$. Then we can write the amount of leakage after an infinitesimal time interval $\delta t$, 
\begin{align}\label{eq:PertLeakageDef}
    l(\delta t) \approx& - \Tr_\| \left[(\mathcal{L}_0 +\mathcal{L}_1 )\hat{\rho} (0)\right ]  \delta t \approx-  \Tr_\| \left[\mathcal{L}_1\hat{\rho}(0)\right ]\delta t.
\end{align} 
In particular, if the perturbation $\mathcal{L}_1=\mathcal{D}[\hat{L}]$ is such that vectors $\hat{L}\vert 0_c\rangle$ and $\hat{L}\vert 1_c\rangle$ lie in the computational subspace, then $\Tr_\| \left\{\mathcal{D}[\hat{L}]\hat{\rho}(0)\right \}=0$ (see Appendix~\ref{app:completelyNonLeaking}).
That is the case for the photon loss, $\Tr_\| \left\{\mathcal{D}[\hat{a}]\hat{\rho}(0)\right \} =0$, so it does not cause leakage in the first order. Therefore, the first-order bit-flip rate is produced by the dynamics in the computational space which is not affected by the Lindbladian $\mathcal{L}_0$. 
It is given by,
\begin{align}
 \Gamma^{(1)}_{0 \leftrightarrow 1} =\frac{1}{2} |\langle1_{c}\vert\hat{a}\vert0_{c}\rangle|^{2}+\frac{1}{2}|\langle0_{c}\vert\hat{a}\vert1_{c}\rangle|^{2} \approx
\kappa_1\alpha^{2}e^{-4\alpha^{2}}
\end{align} 
for $\alpha^2\gg 1$~\cite{Gautier2022}. 
However, in the second order the photon loss causes leakage,  $\Tr_\| \left[(\mathcal{D}[\hat{a}])^2\hat{\rho}(0)\right ] \neq 0 $. This condition emerges expanding $l(\delta t)$ to the second order in $\delta t$,
\begin{align}\label{eq:PertLeakageDef2}
    l(\delta t) =&-  \Tr_\| \left\{[\delta t(\mathcal{L}_0+\mathcal{L}_1)+\frac{\delta t^2}{2}(\mathcal{L}_0+\mathcal{L}_1)^2]\hat{\rho}(0)\right\}\nonumber\\
    &=- \Tr_\| \left[(\mathcal{L}_1)^2\hat{\rho}(0)\right] \frac{\delta t^2}{2},
\end{align}
where we used that if the perturbation $\mathcal{L}_1$ does not cause leakage in the first order, only the term $(\mathcal{L}_1)^2$ survives (see Appendix~\ref{app:completelyNonLeaking} for details). Therefore, to treat the photon loss, one has to include the non-computational states, through which the probability flows. Using our calculation scheme, we find the full bit-flip rate with the leading exponential $e^{-2\alpha^2}$ (see Table~\ref{tab:bit-flip}).

Other perturbations for which the first-order perturbative analysis does not yield the decay rates are Hamiltonian perturbations of the form $\mathcal{L}_1 = -i[\hat{V},\hat{\rho}]$. Examples are an uncompensated frequency shift (detuning)~\cite{Lescanne2020} and gate implementation. Because first-order contributions to the computational subspace eigenvalues are purely imaginary, to obtain the decay rates, one needs to find real second-order contributions. 
Following our calculation scheme, it is possible to find the second-order decay rates of the cat qubit due to, hypothetically, any Hamiltonian type perturbation. We perform calculations of the bit-flip rate induced by the $Z$ gate $\mathcal{L}_1=-i\epsilon_Z [\hat{a}^\dagger + \hat{a}, \hat{\rho}]$ and by the detuning $\mathcal{L}_1=-i\Delta[\hat{a}^\dagger \hat{a}, \hat{\rho}]$ in Appendixes~\ref{app:Zgate} and~\ref{app:detuning}.
In Table~\ref{tab:bit-flip} we present just the results. 
\begin{table}
    \centering
    \begin{tabular}{|c|c|c|}
        \hline
        Photon loss & $Z$ gate & Detuning  \\
        \hline
         $\displaystyle \kappa_1\alpha^2 e^{-4\alpha^2} + \frac{\kappa_{1}^{2}}{2\kappa_{2}}e^{-2\alpha^{2}}$ & $\displaystyle \frac{2 \epsilon_Z^2}{\kappa_{2}\alpha^{2}}e^{-2\alpha^{2}} $  & $ \displaystyle 2\frac{\Delta^2}{\kappa_2} e^{-2\alpha^2}$ \\
         \hline
    \end{tabular}
    \caption{Contributions to the bit-flip error rate ($X$ error) from the single-photon loss, the $Z$ gate, and the detuning. For the bit-flip rate due to the photon loss for the cat-state size $\alpha^2\gtrsim \ln(\kappa_2/\kappa_1)/2$, the second-order contribution dominates. Note that the effect of the detuning perturbation was studied in the theoretical work~\cite{Thompson2022} by instanton technique. In particular, there it was found that the bit-flip rate is proportional to $\propto (\Delta^2/\kappa_2)e^{-2\alpha^2}$, but the numerical prefactor was not derived.}
    \label{tab:bit-flip}
\end{table}

\section{Second-order perturbation theory for the cat qubit}
To develop perturbation theory for a generic Lindbladian $\mathcal{L}_0$ with a perturbation $\mathcal{L}_1$, it is convenient to introduce two projectors. The first one $\mathcal{P}_\|\equiv\lim_{t\to\infty}e^{\mathcal{L}_0 t}$ projects onto the stable degenerate subspace (the cat subspace in our case), such that $\mathcal{L}_0\mathcal{P}_\|\hat{\rho}=0$ for any density matrix $\hat{\rho}$. The other projector $\mathcal{P}_\|=1-\mathcal{P}_\perp$ is such that $\lim_{t\to\infty}e^{\mathcal{L}_0 t}\mathcal{P}_\perp\hat{\rho}=0$ for any density matrix $\hat{\rho}$ or, equivalently, $\mathcal{L}_0^{-1}$ is defined for $\mathcal{P}_\perp\hat{\rho}$. Then defining $\hat{\rho}_\| \equiv \mathcal{P}_\| \hat{\rho}$ and $\hat{\rho}_\perp \equiv \mathcal{P}_\perp \hat{\rho}$, the master equation with the perturbation $\mathcal{L}_1$ for the components $\hat{\rho}_\|$ and $\hat{\rho}_\perp$ becomes
\begin{subequations}
    \begin{align}
    \frac{d\hat{\rho}_\|}{dt}&=\mathcal{P}_\|\mathcal{L}_1 \hat{\rho}_\| + \mathcal{P}_\| \mathcal{L}_1 \hat{\rho}_\perp, \\
    \frac{d\hat{\rho}_\perp}{dt}&=\mathcal{L}_0 \hat{\rho}_\perp + \mathcal{P}_\perp \mathcal{L}_1 \hat{\rho}_\perp + \mathcal{P}_\perp \mathcal{L}_1 \hat{\rho}_\|.
    \end{align}
\end{subequations}
The component $\hat{\rho}_\perp$ decays fast with the time scale given by $\mathcal{L}_0$. Therefore, assuming $d\hat{\rho}_\perp / dt = 0$, we may iteratively solve the master equation for it and obtain $\hat{\rho}_\perp=-\mathcal{L}_0^{-1}\mathcal{P}_\perp \mathcal{L}_1 \hat{\rho}_\| + O(\mathcal{L}_1^2)$. Substituting this into the master equation for the slow component $\hat{\rho}_\|$, we find
\begin{equation}\label{eq:masterprojected}
\frac{d\hat\rho_\|}{d t }=\mathcal{P}_\|\mathcal{L}_1\hat\rho_\|
-\mathcal{P}_\|\mathcal{L}_1\mathcal{L}_0^{-1}\mathcal{P}_\perp\mathcal{L}_1\hat\rho_\| ,
\end{equation}
where we neglect terms of $O(\mathcal{L}_1^3)$  and higher (see Ref.~\cite{Regent2023} for a systematic treatment of higher orders).
On the right-hand side (rhs) of Eq.~\eqref{eq:masterprojected} we have superoperators which are basically $4\times4$ matrices in the computational subspace. The first superoperator yields the error rates in the first order. As we explained before, if the perturbation $\mathcal{L}_1$ does not cause leakage, this superoperator does not capture the probability transfer mechanism through the noncomputational subspace. Therefore, for such perturbations it is necessary to study the second-order perturbation term, which is the second superoperator on the rhs of Eq.~\eqref{eq:masterprojected}. Calculation of the second-order term for perturbations, which do not cause leakage in the first order, is our goal now.   

To study the rhs of Eq.~\eqref{eq:masterprojected} we need to be able to do two things. First is to project on the cat subspace, that is, to act with the projector $\mathcal{P}_\|$. This is done with the help of the left eigenoperators of $\mathcal{L}_0$, so-called invariants of the dynamics~\cite{Mirrahimi2014}:\begin{subequations}\label{eqs:lefteigenvectors}
\begin{align}
&\hat\varsigma_0^{++}=\sum_{n=0}^\infty|2n\rangle\langle2n|,\\
&\hat\varsigma_0^{--}=\sum_{n=0}^\infty|2n+1\rangle\langle2n+1|,\\
&\hat\varsigma_0^{+-}=\mathcal{N}_0
\sum_{n,m=0}^\infty\frac{(-1)^{n-m}I_{n-m}(\alpha^2)}{2n+1-2m}\,
\frac{\hat{a}^\dagger{}^{2n+1}|0\rangle\langle0|\hat{a}^{2m}}{(2n)!!\,(2m)!!}\nonumber\\
&\qquad{}=(\hat\varsigma_0^{-+})^\dagger,
\end{align}\end{subequations}
where $I_{n-m}(z)$ is the modified Bessel function, $\vert n \rangle$ is an $n$-photon Fock state, and $\mathcal{N}_0=\sqrt{\frac{2\alpha^2}{\sinh2\alpha^2}}$ is a normalization factor. In particular, these invariants of the dynamics provide an equivalent form of the superoperator $\mathcal{P}_\| \hat{\rho}= \sum_{\sigma \sigma^\prime}\vert \mathcal{C}_\alpha^{\sigma}\rangle \langle \mathcal{C}_\alpha^{\sigma^\prime} \vert \Tr(\hat{\varsigma}^{\sigma^\prime \sigma} \hat{\rho}) $. Therefore, they are enough to perform first-order calculations, but for the second-order treatment we also need to be able to act with $\mathcal{L}_0^{-1}$. In principle, this requires knowledge of the complete eigensystem of $\mathcal{L}_0$. However, for the perturbations, that do not cause leakage in the first order, the superoperator action $\mathcal{L}_1$ on the computational state has the operator structure, $\mathcal{L}_1\hat\rho_\| =  \vert v_{1}\rangle\langle\mathcal{C}_{\alpha}^{+}\vert+\vert v_{2}\rangle\langle\mathcal{C}_{\alpha}^{-}\vert+\text{h.c.}$, where $\vert v_{1,2}\rangle$ are vectors in the oscillator's Hilbert space. Indeed, if $\mathcal{L}_1\hat{\rho}_\| = -i[\hat{V},\hat{\rho}_\|]$ is of the Hamiltonian type, then for each term in $\hat{\rho}_\|$ the action changes either  the bra or ket part of the term, but not both simultaneously. If $\mathcal{L}_1\hat{\rho}_\| = \mathcal{D}[\hat{L}]\hat{\rho}_\|$, then $\mathcal{L}_1$ not causing leakage in the first order means that the vector $\hat{L}\vert\mathcal{C_\alpha^\pm}\rangle $ is a linear combination of the cat states (see Appendix~\ref{app:completelyNonLeaking}). Evidently, $\hat{L}\hat{\rho}_\|\hat{L}^\dagger$ belongs to the computational subspace, while the anticommutator part of $ \mathcal{D}[\hat{L}]$ follows the same logic as the Hamiltonian-type perturbation. 

Second, to act with $\mathcal{L}_0^{-1}$, we introduce an auxiliary basis, that of the Kerr Hamiltonian eigenstates, defined by
\begin{equation}
\label{eq:KerrHamiltonian}
\hat{H}|\psi_{l\sigma}\rangle\equiv(\hat{a}^\dagger{}^2-\alpha^2)(\hat{a}^2-\alpha^2)|\psi_{l\sigma}\rangle=\mu_{l\sigma}|\psi_{l\sigma}\rangle.
\end{equation}
Here $\alpha$ is a positive number, $\sigma$ denotes the parity of the eigenvectors $|\psi_{l\sigma}\rangle$ in the sense of $(-1)^{\hat{a}^\dagger\hat{a}} |\psi_{l\sigma}\rangle = \sigma |\psi_{l\sigma}\rangle$ and the integer $l\geq0$ is a quantum number. In particular, the cat states are $|\mathcal{C}_\alpha^{\sigma}\rangle = |\psi_{0\sigma}\rangle$.
Then it is straightforward to see that
\begin{subequations}\label{eqs:Lindbladian_eigenvectorsl00l}
\begin{align}
&\mathcal{L}_0|\psi_{l\sigma}\rangle\langle\mathcal{C}_\alpha^{\sigma'}| =
-\frac{\kappa_2\mu_{l\sigma}}2\,|\psi_{l\sigma}\rangle\langle\mathcal{C}_\alpha^{\sigma'}|,\\
&\mathcal{L}_0|\mathcal{C}_\alpha^{\sigma}\rangle\langle\psi_{l'\sigma'}| =
-\frac{\kappa_2\mu_{l'\sigma'}}2\,|\mathcal{C}_\alpha^{\sigma}\rangle\langle\psi_{l'\sigma'}|. 
\end{align}\end{subequations}
It is clear that these right eigenoperators are enough to expand the action of a perturbation, which does not cause leakage in the first order, on the computational state $\mathcal{L}_1\hat\rho_\| =  \vert v_{1}\rangle\langle\mathcal{C}_{\alpha}^{+}\vert+\vert v_{2}\rangle\langle\mathcal{C}_{\alpha}^{-}\vert+\text{h.c.}$ in the linear combination of the right eigenvectors of $\mathcal{L}_0$. Indeed, inserting the resolution of identity in terms of the Kerr eigenstates $\hat{I}=\sum_{l\sigma}\vert\psi_{l\sigma}\rangle\langle\psi_{l\sigma}\vert$, we form the right eigenoperators~\eqref{eqs:Lindbladian_eigenvectorsl00l}
\begin{align}\label{eq:genericaction}
    \mathcal{L}_1\hat\rho_\| =& \sum_{l,\sigma} \left(\langle\psi_{l\sigma}\vert v_{1}\rangle \vert\psi_{l\sigma}\rangle\langle\mathcal{C}_{\alpha}^{+}\vert+\langle\psi_{l\sigma}\vert v_{2}\rangle \vert\psi_{l\sigma}\rangle\langle\mathcal{C}_{\alpha}^{-}\vert\right)\nonumber\\
    &+\text{H.c.}
\end{align}
Then to act with the projectors $\mathcal{P_\|}$ and $\mathcal{P_\perp}$ on the $\mathcal{L}_1\hat\rho_\|$ we pick either terms with $l=0$ or terms with $l\geq 1$ in Eq.~\eqref{eq:genericaction}. Formally, $\mathcal{P}_\|$ erases the right eigenvectors~\eqref{eqs:Lindbladian_eigenvectorsl00l} with $l>0$, because they have negative eigenvalues. Also, such expansion in the right eigenvectors trivializes action with the inverse $\mathcal{L}_0$: each term is simply divided by the corresponding eigenvalue.  Note that if the perturbation $\mathcal{L}_1$ causes leakage in the first order, then its action on the computational subspace $\mathcal{L}_1\hat\rho_\|$ contains terms of the form $\vert\psi_{l\sigma}\rangle\langle\psi_{l\sigma}\vert$ with $l\geq 1$.  Then the right eigenvectors~\eqref{eqs:Lindbladian_eigenvectorsl00l} are not enough to expand the action in that case, but it is not needed: The leading exponential $e^{-2\alpha^2}$ in the bit-flip rate should be found in the first order.\footnote{We do not to prove this statement for a generic leaking perturbation, but we do the calculation for some of the physically relevant perturbations (see Appendix~\ref{app:leakingperturbations}).} 

Now we are ready to begin the calculations of the rhs of Eq.~\eqref{eq:masterprojected}. As noted before, in this paper we derive the bit-flip rate induced by the photon loss $\mathcal{L}_{1}=\kappa_1\mathcal{D}\left[\hat{a}\right]$ in the second order in $\kappa_1$. The main steps of the calculation scheme are presented in the main text, while technical details are shown in Appendix~\ref{app:photonloss}. The bit-flip rate is essentially the decay rate of the $z$ component of the density matrix in the Bloch representation. Thus, for the second-order correction to the bit-flip rate we have 
\begin{equation}\label{eq:bitflip0}
\Gamma^{(2)}_{0\leftrightarrow1}=\frac{\kappa_1^2}{2}\text{Tr}\left(\hat\varsigma_0^{z}\mathcal{D}[\hat{a}]\mathcal{L}_{0}^{-1}\mathcal{P}_{\perp}\mathcal{D}[\hat{a}]\hat{\varrho}_{0}^{z}\right).
\end{equation}
Here, for brevity we defined the $z$ axis operator of the qubit density matrix as $\hat\varrho_0^z=|\mathcal{C}_\alpha^+\rangle\langle\mathcal{C}_\alpha^{-}|+|\mathcal{C}_\alpha^-\rangle\langle\mathcal{C}_\alpha^{+}|$. The projection on this component is done by taking the trace with $\hat\varsigma_0^{z}=(\hat\varsigma_0^{+-}+\hat\varsigma_0^{-+})/2$, which can be seen by recalling the property of the left eigenvectors~\eqref{eqs:lefteigenvectors} $\Tr(\varsigma_0^{\sigma {\sigma^\prime}} |\mathcal{C}_\alpha^{\sigma^\prime_1}\rangle\langle\mathcal{C}_\alpha^{\sigma_1}|) = \delta_{\sigma \sigma_1}\delta_{\sigma^\prime \sigma^\prime_1}$~\cite{Mirrahimi2014}. The calculation of $\Gamma^{(2)}_{0\leftrightarrow1}$ consists of two main steps. First, we get rid of the Lindbladian left eigenoperators and express the bit-flip rate only via the Kerr Hamiltonian eigensystem~\eqref{eq:KerrHamiltonian}, namely, as a certain matrix element of the inverted Kerr Hamiltonian, $\hat{H}_{\perp}^{-1}\equiv \sum_{l>0,\sigma}\frac{\vert\mathcal{\psi}_{l\sigma}\rangle\langle\mathcal{\psi}_{l\sigma}\vert}{\mu_{l\sigma}}$. Second, we explicitly find $\hat{H}_{\perp}^{-1}$, by representing the Kerr Hamiltonian in the basis of coherent states, where it becomes a second-order differential operator. Then, inverting this operator, we obtain the matrix element of $\hat{H}_{\perp}^{-1}$ in the basis of coherent states.

\subsection{From Lindbladian to Kerr Hamiltonian}
We start by transforming the expression for $\Gamma^{(2)}_{0\leftrightarrow1}$~\eqref{eq:bitflip0} such that it contains only the Kerr Hamiltonian eigensystem. We act with $\mathcal{L}_{1}=\kappa_1\mathcal{D}\left[\hat{a}\right]$ on the $z$ axis operator $\hat{\varrho}_0^z$. After that we insert resolutions of identity in terms of the Kerr eigenstates $\hat{I}=\sum_{l\sigma}\vert\psi_{l\sigma}\rangle\langle\psi_{l\sigma}\vert$ to form the right eigenoperators~\eqref{eqs:Lindbladian_eigenvectorsl00l},
\begin{align} 
\mathcal{D}\left[\hat{a}\right]\hat{\varrho}_{0}^z =\alpha^{2}\hat{\varrho}_{0}^z -\frac{1}{2}\!\sum_{l\geq0,\sigma}^{\infty}\!\left(\langle\psi_{l\sigma}\vert\hat{a}^{\dagger}\hat{a}\vert\mathcal{C}_{\alpha}^{\sigma}\rangle\vert\psi_{l\sigma}\rangle\langle\mathcal{C}_{\alpha}^{-\sigma}\vert +\text{H.c.}\right).
\end{align}
Action with the projector $\mathcal{P}_{\perp}$ corresponds to picking the eigenoperators with negative eigenvalue
\begin{equation}\label{eq:PperpDaRhoz}
\mathcal{P}_{\perp}\mathcal{D}[\hat{a}]\hat{\varrho}_{0}^{z}=-\frac{1}{2}\sum_{l\geq1,\sigma}\langle\psi_{l\sigma}\vert\hat{a}^{\dagger}\hat{a}\vert\mathcal{C}_{\alpha}^{\sigma}\rangle\vert\psi_{l\sigma}\rangle\langle\mathcal{C}_{\alpha}^{-\sigma}\vert+\text{H.c.}
\end{equation}
Then, to apply $\mathcal{L}_0^{-1}$ we simply divide the eigenoperators in Eq.~\eqref{eq:PperpDaRhoz} with their eigenvalues~\eqref{eqs:Lindbladian_eigenvectorsl00l}. Finally, we act again with the perturbation $\mathcal{L}_1$ and rearrange the elements under the trace using the cyclic property (such that $\mathcal{D}[\hat{a}]$ acts on the left eigenoperator $\hat\varsigma_0^{z}=(\hat\varsigma_0^{+-}+\hat\varsigma_0^{-+})/2$ from the right). The intermediate result for the bit-flip rate due to the photon loss is
\begin{align}
   \Gamma_{0 \leftrightarrow 1}^{(2)}={}&{}\kappa_{1}^{2}\sum_{l>0,\sigma}\frac{\langle\psi_{l\sigma}\vert\hat{a}^{\dagger}\hat{a}\vert\mathcal{C}_{\alpha}^{\sigma}\rangle}{4\kappa_{2}\mu_{l\sigma}}\Bigl(\langle\mathcal{C}_{\alpha}^{-\sigma}\vert\hat{a}^{\dagger}\hat{\varsigma}_{0}^{-\sigma,\sigma}\hat{a}\vert\psi_{l\sigma}\rangle\nonumber\\
&-\frac{1}{2}\langle\mathcal{C}_{\alpha}^{-\sigma}\vert\left\{ \hat{\varsigma}_{0}^{\sigma,-\sigma},\hat{a}^{\dagger}\hat{a}\right\} \vert\psi_{l\sigma}\rangle\Bigr)+\text{h.c.},
\label{eq:secondorderforloss_intermediate}
\end{align} 
where $\{\hat{A},\hat{B}\}\equiv\hat{A}\hat{B} + \hat{B}\hat{A}$.
The task for now is to evaluate the matrix elements of the left eigenoperators~\eqref{eqs:lefteigenvectors}. 
To evaluate the first matrix element, we note that $\langle\mathcal{C}_{\alpha}^{\sigma}|\hat{\varsigma}_{0}^{\sigma',\sigma}|\psi_{l\sigma'}\rangle = 0$ for $l>0$, since $|\psi_{l\sigma'}\rangle\langle\mathcal{C}_{\alpha}^{\sigma}|$ is a right eigenvector. Thus, inserting a resolution of identity $\hat{I}=\sum_{l\sigma}\vert\psi_{l\sigma}\rangle\langle\psi_{l\sigma}\vert$ after $\hat{\varsigma}_{0}^{-\sigma,\sigma}$, we find
\begin{align} \label{eq:matrixelementoflefteigenvector_main}
     &\langle\mathcal{C}_{\alpha}^{-\sigma}|\hat{a}^{\dagger}\hat{\varsigma}_{0}^{-\sigma,\sigma}\hat{a}|\psi_{l\sigma}\rangle=\alpha\left(\tanh\alpha^{2}\right)^{-\sigma/2}\langle\mathcal{C}_{\alpha}^{-\sigma}|\hat{a}|\psi_{l\sigma}\rangle.
\end{align}
For the term with the anticommutator we will proceed as follows. First, we write it as a sum of the commutator and an extra term. For the extra term the evaluation is done like in \eqref{eq:matrixelementoflefteigenvector_main} and we find
\begin{align}
    &\frac{1}{2}\langle\mathcal{C}_{\alpha}^{-\sigma}|\{\hat{a}^{\dagger}\hat{a},\hat{\varsigma}_{0}^{\sigma,-\sigma}\}|\psi_{l\sigma}\rangle\nonumber\\
    &=\langle\mathcal{C}_{\alpha}^{-\sigma}|\hat{\varsigma}_{0}^{\sigma,-\sigma}\hat{a}^{\dagger}\hat{a}|\psi_{l\sigma}\rangle+\frac{1}{2}\langle\mathcal{C}_{\alpha}^{-\sigma}|\left[\hat{a}^{\dagger}\hat{a},\hat{\varsigma}_{0}^{\sigma,-\sigma}\right]|\psi_{l\sigma}\rangle\nonumber\\
&=\langle\mathcal{C}_{\alpha}^{\sigma}|\hat{a}^{\dagger}\hat{a}|\psi_{l\sigma}\rangle+\frac{1}{2}\langle\mathcal{C}_{\alpha}^{-\sigma}|\left[\hat{a}^{\dagger}\hat{a},\hat{\varsigma}_{0}^{\sigma,-\sigma}\right]|\psi_{l\sigma}\rangle.
\end{align}
To deal with the term with the commutator we have to use the explicit form of $\hat{\varsigma}_{0}^{\sigma,-\sigma}$~\eqref{eqs:lefteigenvectors}. At this point one can see why we transformed the anticommutator into the commutator: The latter cancels the denominator $2n + 1 - 2m$ in the expression for $\hat{\varsigma}_{0}^{+-}$ and $\hat{\varsigma}_{0}^{-+}$, so that the summation over $n$ and $m$ simplifies. We use an integral representation for the modified Bessel function $I_n(z)=\int_{-\pi}^\pi\frac{d\phi}{2\pi}\,e^{z\cos\phi+in\phi}$ to evaluate the sum at the cost of introducing the integration,
\begin{subequations}
\label{eqs:projectionOfNon-comp_main}\begin{align}
  &{}\langle\mathcal{C}_{\alpha}^{-}\vert\left[\hat{a}^{\dagger}\hat{a},\varsigma_{0}^{+-}\right]\vert\mathcal{\psi}_{l+}\rangle\nonumber\\
    {}&{} =\mathcal{N}_0 \sum_{n,m=0}^\infty \frac{(-1)^{n-m}I_{n-m}(\alpha^2)}{(2n)!!\,(2m)!!}\,\frac{\alpha^{2n+1}}{\sqrt{\sinh\alpha^2}}\,\langle0|\hat{a}^{2m}|\psi_{l+}\rangle\nonumber\\
    {}&{} = \mathcal{N}_0\frac{\alpha}{\sqrt{\sinh\alpha^2}}
    \int_{-\pi}^\pi\frac{d\phi}{2\pi}\,
    \langle0|\exp[(\alpha^2-\hat{a}^2)e^{-i\phi}/2]|\psi_{l+}\rangle \nonumber\\
    {}&{} = \mathcal{N}_0\frac{\alpha\,\langle0|\psi_{l+}\rangle}{\sqrt{\sinh\alpha^2}},\\
    {}&{}\langle\mathcal{C}_{\alpha}^+\vert\left[\hat{a}^{\dagger}\hat{a},\varsigma_{0}^{-+}\right]\vert\mathcal{\psi}_{l-}\rangle
    =-\mathcal{N}_0\frac{\langle0|\hat{a}|\psi_{l-}\rangle}{\sqrt{\cosh\alpha^2}},
\end{align}\end{subequations}
where $\mathcal{N}_0=\sqrt{\frac{2\alpha^2}{\sinh2\alpha^2}}$ is a normalization factor.
Substituting these simplified matrix elements back into \eqref{eq:secondorderforloss_intermediate} and recalling the definition of the Schrödinger cat states~\eqref{eq:catstates}, we find 
\begin{align}
\label{eq:bitflip1}
\Gamma_{0\leftrightarrow1}^{(2)}={}&{}\frac{\kappa_{1}^{2}}{2\kappa_{2}}\frac{\alpha^{2}}{\sinh2\alpha^{2}}\Bigl[e^{\alpha^{2}/2}\langle0\vert\left(\hat{a}-\alpha\right)\hat{H}_{\perp}^{-1}\hat{a}^{\dagger}\vert\alpha\rangle\nonumber\\
&-2\langle-\alpha\vert\hat{a}\hat{H}_{\perp}^{-1}\hat{a}^{\dagger}\vert\alpha\rangle\Bigr].
\end{align}
So far, we have managed to express the second-order correction to the bit-flip rate $\Gamma_{0\leftrightarrow1}^{(2)}$ in terms of the Kerr Hamiltonian eigenstates, which is convenient  for both numerical and analytical calculation. Next we will evaluate the matrix elements of the inverted Kerr Hamiltonian in Eq.~\eqref{eq:bitflip1} exactly for any $\alpha > 0$ by inverting the Kerr Hamiltonian in the basis of coherent states.

\subsection{Inverting the Kerr Hamiltonian}
We use the basis of unnormalized coherent states, defined in the Fock basis as $\vert\chi\rangle=\sum_{n=0}^{\infty}\left(\chi^{n}/\sqrt{n!}\right)\vert n\rangle$. Then, the matrix element of the inverse Kerr Hamiltonian  between two arbitrary unnormalized coherent states, $H_{\perp}^{-1}(\bar{\chi},\varphi)\equiv\langle \chi \vert \hat{H}_{\perp}^{-1}\vert\varphi\rangle $ obeys the differential equation 
\begin{align}
    &\left(\bar{\chi}^{2}-\alpha^{2}\right)\left(\partial_{\bar{\chi}}^{2}-\alpha^{2}\right)H_{\perp}^{-1}(\bar{\chi},\varphi)\nonumber\\
    &=\langle\chi\vert(1-\sum_{\sigma}\vert\mathcal{C}_{\alpha}^{\sigma}\rangle\langle\mathcal{C}_{\alpha}^{\sigma}\vert)\vert\varphi\rangle,
    \label{eq:diffeq}
\end{align}
with the boundary conditions $H_{\perp}^{-1}(\pm\alpha,\varphi)=0$. The solution of this differential equation can be expressed via special functions (see Appendix~\ref{app:inverseKerr}). Then the matrix  elements entering the bit-flip rate~\eqref{eq:bitflip1} can be obtained as partial derivatives of $H_{\perp}^{-1}(\bar{\chi},\varphi)$. The final result for the second-order contribution to the bit-flip rate is
\begin{align}
\Gamma_{0\leftrightarrow1}^{(2)}&=\frac{\kappa_{1}^{2}\alpha^{2}\left[\text{Shi}(2\alpha^{2})+\left(1-\coth2\alpha^{2}\right)\text{Chin}(4\alpha^{2})\right]}{2\kappa_{2}\sinh^{2}2\alpha^{2}}\nonumber\\
&\approx\frac{\kappa_{1}^{2}}{2\kappa_{2}}e^{-2\alpha^{2}}\phantom{a}(\alpha \gg 1),
\end{align}
where $\text{Shi}(z)$ and $\text{Chin}(z)$ are hyperbolic sine and cosine\footnote{Our definition of the hyperbolic cosine integral is such that the function is holomorphic in the complex plane and it is related to the standard cosine integral as $\text{Chin}(z)= \text{Chi}(z) - \ln z - \gamma$ with Euler constant $\gamma$.
} integrals~\cite{abramowitz+stegun}.

\section{Discussion}\label{sec:discussion}
Can it happen that the higher-order corrections significantly increase the bit-flip rate? To answer this question let us try to guess the asymptotic behavior at $\alpha \gg 1$ of the $k$-th order perturbation correction  $\lambda_z^{(k)}$ to the eigenvalue $\lambda_z$.

For this, let us assume that the perturbation $\mathcal{L}_1$ of the strength $\kappa_L$ scales as $\kappa_L \alpha^n$ with some power $n$, if we formally replace the creation and annihilation operators $\hat{a}^\dagger, \hat{a}$ with $\alpha$. For example, the single-photon loss scales as $\alpha^2$, Z gate scales as $\alpha$, and the dephasing $\mathcal{D}[\hat{a}^\dagger \hat{a}]$ scales as $\alpha^4$. As we hinted for the exponential $e^{-2\alpha^2}$ to appear, the perturbation should cause leakage in this order. It is important to stress that this exponential is encoded in the structure of the unperturbed Linbladian $\mathcal{L}_0$ as was shown by the nonperturbative instanton calculation~\cite{Thompson2022}.
For simplicity, let us take a perturbation $\mathcal{L}_1 = \mathcal{D}[\hat{L}]$ with $\hat{L} = \hat{a}^{\dagger p} \hat{a}^q$, where the integer powers $p \ge 1, q\geq 0$, $2p+2q=n$, so that $\mathcal{L}_1$ causes leakage in the first order. 

The $k$-th order correction $\lambda_z^{(k)}$ contains, in general, action with $\mathcal{L}_1$ $k$ times, action with $\mathcal{L}_0^{-1}$ $k-1$ times and a mixture of projectors $\mathcal{P}_\|,\mathcal{P}_\perp$.  For example, the third-order contribution to Eq.~\eqref{eq:masterprojected} has the form, 
\begin{equation}
\mathcal{P}_{\|}\mathcal{L}_{1}\left(\mathcal{L}_{0}^{-1}\mathcal{P}_{\perp}\mathcal{L}_{1}\mathcal{L}_{0}^{-1}\mathcal{P}_{\perp}-\mathcal{L}_{0}^{-2}\mathcal{P}_{\perp}\mathcal{L}_{1}\mathcal{P}_{\|}\right)\mathcal{L}_{1}\hat{\rho}_{\|}.
\end{equation} 
Each action with $\mathcal{L}_0^{-1}$ gives a factor $1/(\kappa_2\alpha^2)$, the inverse of the typical non-zero eigenvalues of the unperturbed Linbladian  $\mathcal{L}_0$~\cite{Chamberland2022}. Each action with $\mathcal{L}_1$ gives a factor $\kappa_L\alpha^n$. However, in order to get the leading exponential $e^{-2\alpha^2}$, we have to leave the computational subspace at least once upon the action of $\mathcal{L}_1$, which brings an extra factor $1/\alpha^2$. 
(Indeed, one can estimate, taking for the initial density matrix $\hat{\rho}(0)= \vert0_c\rangle\langle 0_c \vert$ the first order leakage as $l(\delta t)\propto \delta t\,(\langle \alpha \vert \hat{L}^\dagger \hat{L}\vert \alpha \rangle - | \langle \alpha \vert \hat{L}\vert \alpha \rangle |^2)$, in this expression the highest power of $\alpha$ cancels.)

All in all, collecting all the powers of $\alpha$ for the $k$-th order correction (with the leakage) to the bit-flip rate eigenvalue we can conjecture, 
\begin{align}
\lambda_z^{(k)}\sim\frac{(\kappa_L\alpha^n)^k}{(\kappa_2\alpha^2)^{k-1}}\frac{1}{\alpha^2}e^{-2\alpha^2} \sim 
\kappa_2\left(\frac{\kappa_L \alpha^{n-2}}{\kappa_2}\right)^{k}  e^{-2\alpha^2}.
\end{align}
We do not know the numerical coefficients in $\lambda_z^{(k)}$, but assuming they depend only on $k$, we can write for the asymptotic behavior of $\lambda_z$,
\begin{equation}
    \lambda_z\sim\kappa_L \alpha^{n-2} f\left(\frac{\kappa_L \alpha^{n-2}}{\kappa_2}\right)e^{-2\alpha^2},\hspace{1em}\alpha\gg1.
\end{equation}
Here $f(x)$ is a function, that does not depend on the perturbation strength and the cat-state size: $f(x)\sim 1$ for $x\sim 1$. However, it can hypothetically grow for $x\gg1$ (which might be even stronger than exponential), thus significantly increasing the bit-flip rate for $\kappa_L\alpha^{n-2}\gg\kappa_2$. 

This crude analysis shows that for the perturbations with $n\leq 2$, like the Z gate, the single-photon loss, the single-photon gain, the detuning, higher perturbation orders should not break the perturbation theory as long as the perturbation strength $\kappa_L \ll \kappa_2$. However, for the perturbations with $n > 2$, the perturbation theory validity condition is stricter: $\kappa_L\alpha^{n-2}\ll\kappa_2$. Particularly, the dephasing perturbation $\mathcal{L}_1=\kappa_\phi \mathcal{D}[\hat{a}^\dagger \hat{a}]$ scales as $\alpha^4$; thus the first-order correction might significantly deviate from the true bit-flip rate for $\kappa_\phi\alpha^2\sim\kappa_2$. The situation may seem even more dramatic with the perturbations with larger $n>4$. For example such perturbations can be caused by Bogolyubov quasiparticles~\cite{Dubovitskii2024} or interaction with the pump mode~\cite{Carde2024}. However, in the case of quasiparticles, the perturbations arise from the operator $\sin (\hat{\varphi}/2)$, where $\hat{\varphi} \propto \hat{a} + \hat{a}^\dagger$ is the phase drop across a Josephson junction in the qubit circuit. Since $\sin(x)$ is a bounded function in the first place, there is hope that the perturbations arising from it will not spoil the bit-flip suppression, as long as the non-locality of the operator $\sin \hat{\varphi}/2$ remains weak. 

\section{Conclusion}
We developed a perturbation theory in the noise strength on top of the dissipative cat qubit Lindbladian, which allowed us to analytically find first- and second-order contributions to the exponentially suppressed bit-flip rates due to different perturbations. We showed that for perturbations, which do not cause leakage in the first order, but cause it in the second order, the second-order contribution is parametrically larger. The obtained expressions match well the numerical simulations in the whole range of the cat-state size $\alpha^2$. Since our expressions as functions of $\alpha^2$ are exact, they can be used for dissipative cat qubits with intermediate sizes $\alpha^2 \sim 1$.

In the future it could be interesting to see whether our method can be used for other cat qubit challenges. For example, to entangle qubits one requires two qubit gates, which introduce additional errors. During the implementation of two qubit gates some induced error rates are exponentially suppressed with the cat-state size~\cite{Chamberland2022}. These error rates can hardly be studied with the shifted Fock basis approach~\cite{Chamberland2022}, which is based on the perturbative expansion in $1/\alpha$ near the stationary computational states $\vert 0_c \rangle, \vert 1_c \rangle$. Thus, a potential extension of our method to two qubit gates would produce a useful tool for benchmarking errors. 
Another possible application is related to the so-called squeezed cat qubit~\cite{Schlegel2022,Hillmann2023}. This qubit is similar to the dissipative cat qubit, but an additional squeezing of the bosonic qubit mode is applied. An enhancement of the bit-flip suppression is expected if one increases the squeezing parameter $r$. However, in the numerical simulations of the bit-flip rate of the squeezed cat qubit due to the single-photon loss this enhancement becomes prominent only for $r > 0.2$~\cite{Hillmann2023}. The authors suppose that this fact might be related to the interplay between two different exponential dependencies of the single-photon loss bit-flip rate observed in dissipative cat qubits. Therefore it could be interesting to see whether the single-photon loss in the squeezed cat qubit could be studied via the second-order perturbation theory. 

Another open question is the structure of the higher orders of the perturbation theory, which was briefly discussed in Sec.~\ref{sec:discussion}. Is it possible to find the bit-flip rate for the dephasing $\kappa_\phi \mathcal{D}[\hat{a}^\dagger\hat{a}]$ non-perturbatively? In some experiments it is the dominant noise channel~\cite{Marquet2024} and it might invalidate the perturbative result for $\kappa_\phi \alpha^2 \sim \kappa_2$.

\acknowledgments
I am grateful to D. M. Basko, J. S. Meyer and M. Houzet for critical reading of the manuscript. I thank Fabrizio Minganti, Ronan Gautier and Joachim Cohen for fruitful discussion.
The work was funded from the Plan France 2030 through the project ANR-22-PETQ-0006.
\appendix

\section{Bit-flip rate of leaking perturbations}
\label{app:leakingperturbations}

In the main text we stated that for leaking perturbations the first-order contribution to the bit-flip rate is $\propto e^{-2\alpha^2}$. Here we justify this statement by calculating the first-order contributions to the perturbations of the general form $\mathcal{L}_1 = \mathcal{D}[a^{\dagger m} a^n]$ with non-negative integers $m,n$. Some of the perturbations, like the photon gain $\mathcal{L}_1 =\mathcal{D}[a^{\dagger}]$ or the dephasing $\mathcal{L}_1 =\mathcal{D}[a^{\dagger }\hat{a}]$ were studied before~\cite{Mirrahimi2014,Dubovitskii2024}. First-order correction to the eigenvalue of the full Lindbladian corresponding to the $z$ component of the Bloch vector is given by, 
\begin{align}
    \lambda_z^{(1)} &=\text{Tr } \frac{(\hat{\varsigma}_0^{+-}+\hat{\varsigma}_0^{-+} )}{2}\mathcal{L}_{1}\left(\vert0_{c}\rangle\langle0_{c}\vert-\vert1_{c}\rangle\langle1_{c}\vert\right) \nonumber \\
    &= -\langle0_{c}\vert\hat{L}^{\dagger}\hat{\varsigma}_{11}\hat{L}\vert0_{c}\rangle-\langle1_{c}\vert\hat{L}^{\dagger}\hat{\varsigma}_{00}\hat{L}\vert1_{c}\rangle,
\end{align}
where $\hat{L} = a^{\dagger m} a^n$ and $\hat{\varsigma}_{00}, \hat{\varsigma}_{11}=1/2\pm(\hat{\varsigma}_0^{+-}+\hat{\varsigma}_0^{-+})/2$ are left eigenvectors corresponding to the computational states. Systematic treatment of such matrix elements of the left eigenvectors $\hat{\varsigma}_0^{+-},\hat{\varsigma}_0^{-+}$ can be performed in the basis of unnormalized coherent states. Starting from the explicit form of $\hat{\varsigma}_0^{+-}$~\eqref{eqs:lefteigenvectors}, using the integral representation for the modified Bessel function $I_n(z)=\int_{-\pi}^\pi\frac{d\phi}{2\pi}\,e^{z\cos\phi+in\phi}$ and an identity $1/(2n-2m+1)=(i/2)\int_0^\pi{e}^{-(2n-2m+1)i\theta}\,d\theta$ we find, 
\begin{align}\label{eq:lefteigenvectorcoherent}
&\langle\chi\vert\varsigma_{0}^{+-}\vert\varphi\rangle \nonumber\\
&=\frac{-i\mathcal{N}_{0}\bar{\chi}}{2}\intop_{0}^{\pi}d\theta e^{i\theta}I_{0}\left(\sqrt{\left(\alpha^{2}-\varphi^{2}e^{-2i\theta}\right)\left(\alpha^{2}-\bar{\chi}^{2}e^{2i\theta}\right)}\right).
\end{align}
Then, omitting terms $O(e^{-4\alpha^2})$ we can approximately write, 
\begin{align}
    \lambda_z^{(1)}= -\alpha^{2 n}(\langle\alpha\vert\hat{a}^{m}\hat{a}^{\dagger m}\vert\alpha\rangle-2\langle\alpha\vert\hat{a}^{m}\varsigma_{0}^{+-}\hat{a}^{\dagger m}\vert\alpha\rangle),
\end{align}
where we used the property of the left eigenvector $\varsigma_{0}^{+-}$, $\langle\alpha\vert\hat{a}^{m}\varsigma_{0}^{+-}\hat{a}^{\dagger m}\vert\alpha\rangle=-\langle{-\alpha}\vert\hat{a}^{m}\varsigma_{0}^{+-}\hat{a}^{\dagger m}\vert{-\alpha}\rangle$ and the fact that the matrix elements above are real-valued.
Now, the calculation reduces to differentiation of $\langle\chi\vert\varsigma_{0}^{+-}\vert\varphi\rangle$ with respect to $\varphi$ $m$-times and with respect to $\bar{\chi}$ $m$-times, substituting $\varphi = \chi = \alpha$ and, finally, integrating over $d\theta$. The results for the perturbations with $m\leq 5$ are presented in Table~\ref{tab:firstorderdissipators}.
\begin{table}
    \centering
    \begin{tabular}{|c|c|}
        \hline
        $m$ &  $\Gamma_{0\leftrightarrow 1}^{(1)}$ \\
        \hline
         $1$ & $\alpha^{2 n}e^{-2\alpha^2}$  \\
         \hline
         $2$ & $2\alpha^{2 n+2}e^{-2\alpha^2} $  \\
         \hline
         $3$ & $3\alpha^{2 n}\left(\alpha^{4}+2\right)e^{-2\alpha^2}$  \\
         \hline
          $4$ & $2\alpha^{2 n+2}\left(2\alpha^{4}+27\right)e^{-2\alpha^2}$  \\
         \hline
          $5$ & $5\alpha^{2 n}\left(\alpha^{8}+48\alpha^{4}+24\right)e^{-2\alpha^2}$  \\
         \hline
    \end{tabular}
    \caption{First-order bit-flip rates induced by the perturbations $\mathcal{D}[a^{\dagger m} a^n]$. The leading exponential is present, while the prefactor is polynomial of $\alpha^2$ of the degree $n+m-1$.}
    \label{tab:firstorderdissipators}
\end{table}
We see that, indeed, the leading exponential is found in the first order, while the prefactor is a polynomial of $\alpha^2$. The highest power term in this polynomial is $m \alpha^{2(n + m - 1)}$.

\section{Completely non-leaking perturbations}\label{app:completelyNonLeaking}

In the main text we claimed that if a perturbation $\mathcal{L}_1$ does not cause leakage in the first order nor in the second order, such a perturbation does not cause leakage in any order, so it is completely non-leaking. 

Let us start with dissipator-type perturbation $\mathcal{D}[\hat{L}]$ produced by a jump operator $\hat{L}$. From the assumption that the perturbation does not cause leakage in the first order, it follows that
\begin{equation}\label{eq:first_order_nonleaking_condition_app}
    \Tr_\| \left[\mathcal{D}[\hat{L}]\hat{\rho}(0)\right ] = 0.
\end{equation}
One should take for the initial density matrix $\hat{\rho}(0)$ pure states $\vert0_c \rangle  \langle 0_c\vert$ and  $\vert1_c \rangle  \langle 1_c\vert$, because the bit-flip rate $\Gamma_{0 \leftrightarrow 1}$ is determined by the transitions from the state $\vert 0_c\rangle$ to the state  $\vert 1_c\rangle$ and vice versa. 

We now show that from the condition~\eqref{eq:first_order_nonleaking_condition_app} it follows that the vectors $\hat{L}\vert 0_c\rangle$ and $\hat{L}\vert 1_c\rangle$ lie in the computational subspace.
To see that we fix some orthonormal basis $\{\vert \psi_n\rangle \}$ in the Hilbert space of the bosonic mode $\hat{a}$, such that $\vert \psi_0\rangle \equiv \vert 0_c\rangle$, $\vert \psi_1\rangle \equiv \vert 1_c\rangle$. From the condition~\eqref{eq:first_order_nonleaking_condition_app} with $\hat{\rho}(0) = \vert0_c \rangle  \langle 0_c\vert$ and the fact that the action $\mathcal{D}[\hat{L}]\hat{\rho}(0)$ is always traceless it follows that
\begin{align}\label{eq:tracePerp}
    0 =& \sum_{n\geq2}\langle \psi_n \vert \mathcal{D}[\hat{L}]\hat{\rho}(0) \vert \psi_n\rangle = \sum_{n\geq2}\langle \psi_n \vert \hat{L}\hat{\rho}(0)\hat{L}^\dagger \vert \psi_n\rangle =\nonumber \\ 
    & \sum_{n\geq2}\vert \langle \psi_n \vert \hat{L} \vert 0_c\rangle \vert^2, 
\end{align}
which can be seen as a trace over non-computational subspace. So the elements $(\hat{L})_{n0} = 0$ for $n \geq 0$ and if we take $\hat{\rho}(0) = \vert1_c \rangle  \langle 1_c\vert$, the elements $(\hat{L})_{n1} = 0$  too. This proves that $\hat{L}\vert 0_c\rangle, \hat{L}\vert 1_c\rangle$ lie in the computational subspace. 

From Eq.~\eqref{eq:tracePerp} it also follows that the converse statement is true. Indeed, if $\hat{L}\vert 0_c\rangle$ and $\hat{L}\vert 1_c\rangle$ lie in the computational subspace, then $\sum_{n\geq2}\langle \psi_n \vert \mathcal{D}[\hat{L}]\hat{\rho}(0) \vert \psi_n\rangle=0$. Since the full trace $\Tr \left[\mathcal{D}[\hat{L}]\hat{\rho}(0)\right ] = 0$ always, we obtain Eq.~\eqref{eq:first_order_nonleaking_condition_app}.

Leakage in the second order is determined by the term $\Tr_\| \left[(\mathcal{D}[\hat{L}])^2\hat{\rho}(0)\right ]$.
Recall that this condition is obtained by expanding the leakage $l(\delta t)$ to the second order in $\delta t$,\begin{align}\label{eq:PertLeakageDef2App}
    &l(\delta t) =-  \Tr_\| \left\{[\delta t(\mathcal{L}_0+\mathcal{L}_1)+\frac{\delta t^2}{2}(\mathcal{L}_0+\mathcal{L}_1)^2]\hat{\rho}(0)\right\}\nonumber\\
    &=- \Tr_\| \left\{[\mathcal{L}_0\mathcal{L}_1+\mathcal{L}_1\mathcal{L}_0+(\mathcal{L}_1)^2]\hat{\rho}(0)\right\} \frac{\delta t^2}{2}.
\end{align}
So far we have dropped the term of first order in $\delta t$, since according to our assumption $\mathcal{L}_1$ does not cause leakage in the first order and obviously $(\mathcal{L}_0)^2$ does not cause leakage. Regarding the cross-terms $\mathcal{L}_0\mathcal{L}_1$ and $\mathcal{L}_1\mathcal{L}_0$, if we act with $\mathcal{L}_0$ first, then we still reside in the computational subspace and since $\mathcal{L}_1$ does not cause leakage in the first order, we have $\Tr_\| \left[\mathcal{L}_1\mathcal{L}_0\hat{\rho}(0)\right] =0$. Next, we act with $\mathcal{L}_1$ first, then as we showed in the main text, its action assumes the form, 
$\mathcal{L}_1\hat\rho(0) =  \vert v_{1}\rangle\langle\mathcal{C}_{\alpha}^{+}\vert+\vert v_{2}\rangle\langle\mathcal{C}_{\alpha}^{-}\vert+\text{h.c.}$. Action with $\mathcal{L}_0$ conserves this form (it only changes the vectors $\vert v_{1,2}\rangle$). Using this form, we can write
\begin{align}
    &\Tr_\| \left[\mathcal{L}_0\mathcal{L}_1\hat{\rho}(0)\right] = -\sum_{n\geq2}\langle \psi_n \vert \mathcal{L}_0\mathcal{L}_1\hat{\rho}(0) \vert \psi_n\rangle\nonumber\\
    ={}&{}-\sum_{n\geq2}\langle \psi_n \vert (\vert v_{1}\rangle\langle\mathcal{C}_{\alpha}^{+}\vert+\vert v_{2}\rangle\langle\mathcal{C}_{\alpha}^{-}\vert+\text{h.c.} )\vert \psi_n\rangle=0,
\end{align}
so the cross term $\mathcal{L}_0\mathcal{L}_1$ also does not cause leakage. This explains that in Eq.~\eqref{eq:PertLeakageDef2App} only the term $\Tr_\| \left[(\mathcal{D}[\hat{L}])^2\hat{\rho}(0)\right ]$ survives for a perturbation $\mathcal{L}_1$, which does not cause leakage in the first order. 

Now we return to our initial assumption that $\mathcal{L}_1$ does not cause leakage in the first order, nor in the second order. The absence of leakage in the second order implies that $\Tr_\| \left[(\mathcal{D}[\hat{L}])^2\hat{\rho}(0)\right ]=0$.
Proceeding in the manner of Eq.~\eqref{eq:tracePerp} we conclude taking $\hat{\rho}(0) = \vert0_c \rangle  \langle 0_c\vert$, 
\begin{align}
    &0 = \sum_{n\geq2}\langle \psi_n \vert (\mathcal{D}[\hat{L}])^2\hat{\rho}(0) \vert \psi_n\rangle = \nonumber \\ &\frac{1}{2}\sum_{n\geq2}\langle \psi_n \vert \hat{L}^\dagger\hat{L}\hat{\rho}(0)\hat{L}^\dagger\hat{L} \vert \psi_n\rangle =
    \sum_{n\geq2}\vert \langle \psi_n \vert\hat{L}^\dagger \hat{L} \vert 0_c\rangle \vert^2.
\end{align}
This proves that the vector $\hat{L}^\dagger\hat{L}\vert 0_c\rangle$ lies in the computational subspace. Starting from $\hat{\rho}(0) = \vert1_c \rangle  \langle 1_c\vert$, we find that $\hat{L}^\dagger\hat{L}\vert 1_c\rangle$ also lies in the computational subspace. Since operators $\hat{L}$ and $\hat{L}^\dagger\hat{L}$ are the only two operators present in the dissipator structure $\mathcal{D}[\hat{L}]$ ($\hat{L}^\dagger$ always acts from the right which is equivalent to $\hat{L}$ acting from the left, which we have analyzed) and they can not produce non-computational amplitudes acting on the computational vector, we conclude that such a jump operator $\hat{L}$ forms completely non-leaking dissipator. 

Finally, we discuss Hamiltonian-type perturbations $-i[\hat{V},\hat{\rho}]$. Due to their structure they never cause leakage in the first order. In the second order one can notice that $(-i[\hat{V},\bullet])^2 = 2\mathcal{D}[\hat{V}]$. So, if a Hamiltonian type perturbation does not cause leakage in the second order, then vectors $\hat{V}\vert0_c \rangle$ and $ \hat{V}\vert1_c \rangle$ belong to the computational subspace. Since $\hat{V}$ is a Hermitian operator, it has to have block-diagonal structure with the computational and the non-computational blocks being separated. Evidently, such operator can not cause leakage in any order. 
As we said, such completely non-leaking perturbations are trivial, because diagonalization of their action truncated to the computational subspace (basically $4 \times 4$ matrix) gives the exact dynamics.

\section{Single photon loss}
\label{app:photonloss}
In this appendix we derive first- and second-order contributions to the bit-flip rate induced by the single photon loss.  Throughout the appendices we use the following notation for the Bloch sphere axis operators of a dissipative cat qubit, 
\begin{subequations} 
\label{eqs:cat_right_eigenvectors}
    \begin{align}
        \hat\varrho_0^I&\equiv|\mathcal{C}_\alpha^+\rangle\langle\mathcal{C}_\alpha^{+}|+|\mathcal{C}_\alpha^-\rangle\langle\mathcal{C}_\alpha^{-}|,\\
        \hat\varrho_0^x&\equiv|\mathcal{C}_\alpha^+\rangle\langle\mathcal{C}_\alpha^{+}|-|\mathcal{C}_\alpha^-\rangle\langle\mathcal{C}_\alpha^{-}|,\\
        \hat\varrho_0^y&\equiv-i|\mathcal{C}_\alpha^+\rangle\langle\mathcal{C}_\alpha^{-}|+i|\mathcal{C}_\alpha^-\rangle\langle\mathcal{C}_\alpha^{+}|,\\
        \hat\varrho_0^z&\equiv|\mathcal{C}_\alpha^+\rangle\langle\mathcal{C}_\alpha^{-}|+|\mathcal{C}_\alpha^-\rangle\langle\mathcal{C}_\alpha^{+}|.
   \end{align}
\end{subequations}

\subsection{First order}
We start with the derivation of the first-order contributions to the cat subspace eigenvalues due to single photon loss. As done in usual degenerate perturbation theory, one needs to project the perturbation onto the degenerate subspace and diagonalize the projected superoperator, that is, the first term in Eq. \eqref{eq:masterprojected}. To find the matrix elements of this operator, we write down the action of $\hat{\mathcal{L}}_{1}=\kappa_1\mathcal{D}\left[\hat{a}\right]$ on the right eigenvectors of the cat space $\hat\varrho_0^{\sigma\sigma'} = \vert \mathcal{C}_\alpha^{\sigma}\rangle \langle \mathcal{C}_\alpha^{\sigma^\prime}\vert$ and insert resolutions of unity in terms of the Kerr eigenstates $\hat{I}=\sum_{l\sigma}\vert\psi_{l\sigma}\rangle\langle\psi_{l\sigma}\vert$ to simplify the projection, 
\begin{align} 
\label{eq:actionofD[a]}
\mathcal{D}\left[\hat{a}\right]\hat{\varrho}_{0}^{\sigma\sigma'} = {}&{} \alpha^{2}\left(\tanh\alpha^{2}\right)^{\frac{\sigma+\sigma^{\prime}}{2}}\vert\mathcal{C}_{\alpha}^{-\sigma}\rangle\langle\mathcal{C}_{\alpha}^{-\sigma'}\vert \nonumber\\&-\sum_{l=0}^{\infty}\frac{\langle\mathcal{C}_{\alpha}^{\sigma'}\vert\hat{a}^{\dagger}\hat{a}\vert\psi_{l\sigma'}\rangle}{2}\vert\mathcal{C}_{\alpha}^{\sigma}\rangle\langle\psi_{l\sigma'}\vert\nonumber\\&-\sum_{l=0}^{\infty}\frac{\langle\psi_{l\sigma}\vert\hat{a}^{\dagger}\hat{a}\vert\mathcal{C}_{\alpha}^{\sigma}\rangle}{2}\vert\psi_{l\sigma}\rangle\langle\mathcal{C}_{\alpha}^{\sigma'}\vert.
\end{align}
Note that the rhs of this expression above is an expansion in the right eigenvectors \eqref{eqs:Lindbladian_eigenvectorsl00l} of the dissipative Lindbladian $\mathcal{L}_0$. Thus the projection onto the cat subspace is merely done by picking $l=0$ terms in the sums over $l$. The resulting 4 by 4 matrix has the form, 
\begin{align}
&\mathcal{P}_{\|}\mathcal{D}[\hat{a}]\vert\mathcal{C}_{\alpha}^{\sigma}\rangle\langle\mathcal{C}_{\alpha}^{\sigma'}\vert=\alpha^{2}\left(\tanh\alpha^{2}\right)^{(\sigma+\sigma')/2}\vert\mathcal{C}_{\alpha}^{-\sigma}\rangle\langle\mathcal{C}_{\alpha}^{-\sigma'}\vert \nonumber\\&-\frac{1}{2}\alpha^2 \left[\left(\tanh\alpha^2\right)^\sigma+\left(\tanh\alpha^2\right)^{\sigma'}\right]\vert\mathcal{C}_{\alpha}^{\sigma}\rangle\langle\mathcal{C}_{\alpha}^{\sigma'}\vert.
\end{align}
It further splits into $2\times2$ blocks with a conserved product of parities $\sigma\sigma'$. Indeed, $\mathcal{D}[\hat{a}]$ either changes the parities of the bra and ket simultaneously or does not change any of them. Proceeding with the diagonalization, we find first-order eigenvalues (the decay rates of the components of the qubit Bloch vector). One of them is $0$, which represents density matrix trace conservation. One is exponentially small for $\alpha \gg 1$,  $\lambda_z^{(1)} = -\kappa_1\alpha^{2}\left(\coth2\alpha^{2}-1\right)\approx -2\kappa_1\alpha^2 e^{-4\alpha^2}.$ The cat subspace part of the right eigenvector corresponding to it is $ \hat\varrho_0^z = \vert\mathcal{C}_{\alpha}^{+}\rangle\langle\mathcal{C}_{\alpha}^{-}\vert+\vert\mathcal{C}_{\alpha}^{-}\rangle\langle\mathcal{C}_{\alpha}^{+}\vert$. The other two eigenvalues that represent the phase-flip rate behave like $\lambda_x^{(1)}\approx\lambda_y^{(1)}\approx -2\kappa_1\alpha^2$ in the limit of large $\alpha \gg 1$. There is not much practical use of finding second-order contributions to them, even though it is possible. Next we evaluate the second-order correction to the exponentially small eigenvalue $\lambda_z$. 

\subsection{Second order}
For the second-order correction we have to evaluate the second term on the rhs of Eq. \eqref{eq:masterprojected}. Since we are interested in the exponentially small eigenvalue we take $\hat\rho_\| =\hat\varrho_0^z$. To project the action of $\mathcal{L}_1$ to the non-computational subspace, we select $l > 0$ terms in the sums of Eq.~\eqref{eq:actionofD[a]}. Then, with the use of Eqs.~\eqref{eqs:Lindbladian_eigenvectorsl00l} we act with $\mathcal{L}_{0}^{-1}$ on these terms,
\begin{equation}
\mathcal{L}_{0}^{-1}\mathcal{P}_{\perp}\mathcal{D}[\hat{a}]\hat{\varrho}_{0}^{z}=\sum_{l>0,\sigma}\frac{\langle\psi_{l\sigma}\vert\hat{a}^{\dagger}\hat{a}\vert\mathcal{C}_{\alpha}^{\sigma}\rangle}{\kappa_{2}\mu_{l\sigma}}\vert\psi_{l\sigma}\rangle\langle\mathcal{C}_{\alpha}^{-\sigma}\vert+\text{h.c.}.
\end{equation}
Now we have to act again with $\mathcal{L}_1$ and finally project back onto $\hat\varrho_0^z$. The last step is done by taking the scalar product with the left eigenvector corresponding to $\hat\varrho_0^z$, that is $\left(\hat{\varsigma}_{0}^{+-} + \hat{\varsigma}_{0}^{-+} \right)/2$. Eliminating the terms which are evidently zero from the parity considerations, we obtain the following expression
\begin{align}
   \lambda_{z}^{(2)}
   =&-\kappa_{1}^{2}\sum_{l>0,\sigma}\frac{\langle\psi_{l\sigma}\vert\hat{a}^{\dagger}\hat{a}\vert\mathcal{C}_{\alpha}^{\sigma}\rangle}{2\kappa_{2}\mu_{l\sigma}}\Bigl(\langle\mathcal{C}_{\alpha}^{-\sigma}\vert\hat{a}^{\dagger}\hat{\varsigma}_{0}^{-\sigma,\sigma}\hat{a}\vert\psi_{l\sigma}\rangle\nonumber\\
&-\frac{1}{2}\langle\mathcal{C}_{\alpha}^{-\sigma}\vert\left\{ \hat{\varsigma}_{0}^{\sigma,-\sigma},\hat{a}^{\dagger}\hat{a}\right\} \vert\psi_{l\sigma}\rangle\Bigr)+\text{h.c.},
\label{eq:secondorderforloss_intermediateApp}
\end{align}
where $\{\hat{A},\hat{B}\}\equiv\hat{A}\hat{B} + \hat{B}\hat{A}$. To evaluate the first matrix element, we note that $\langle\mathcal{C}_{\alpha}^{\sigma}|\hat{\varsigma}_{0}^{\sigma',\sigma}|\psi_{l\sigma'}\rangle = 0$ for $l>0$, since $|\psi_{l\sigma'}\rangle\langle\mathcal{C}_{\alpha}^{\sigma}|$ is a right eigenvector. Thus, inserting a resolution of identity $\hat{I}=\sum_{l\sigma}\vert\psi_{l\sigma}\rangle\langle\psi_{l\sigma}\vert$ after $\hat{\varsigma}_{0}^{-\sigma,\sigma}$, one finds
\begin{align} \label{eq:matrixelementoflefteigenvector}
     &\langle\mathcal{C}_{\alpha}^{-\sigma}|\hat{a}^{\dagger}\hat{\varsigma}_{0}^{-\sigma,\sigma}\hat{a}|\psi_{l\sigma}\rangle=\alpha\left(\tanh\alpha^{2}\right)^{-\frac{\sigma}{2}}\langle\mathcal{C}_{\alpha}^{-\sigma}|\hat{a}|\psi_{l\sigma}\rangle.
\end{align}
For the term with the anticommutator we will proceed as follows. First, we write it as a sum of the commutator and an extra term. For the extra term the evaluation is done like in \eqref{eq:matrixelementoflefteigenvector} and we find
\begin{align}
    &\frac{1}{2}\langle\mathcal{C}_{\alpha}^{-\sigma}|\{\hat{a}^{\dagger}\hat{a},\hat{\varsigma}_{0}^{\sigma,-\sigma}\}|\psi_{l\sigma}\rangle=\nonumber\\
    &\langle\mathcal{C}_{\alpha}^{-\sigma}|\hat{\varsigma}_{0}^{\sigma,-\sigma}\hat{a}^{\dagger}\hat{a}|\psi_{l\sigma}\rangle+\frac{1}{2}\langle\mathcal{C}_{\alpha}^{-\sigma}|\left[\hat{a}^{\dagger}\hat{a},\hat{\varsigma}_{0}^{\sigma,-\sigma}\right]|\psi_{l\sigma}\rangle=\nonumber\\
&\langle\mathcal{C}_{\alpha}^{\sigma}|\hat{a}^{\dagger}\hat{a}|\psi_{l\sigma}\rangle+\frac{1}{2}\langle\mathcal{C}_{\alpha}^{-\sigma}|\left[\hat{a}^{\dagger}\hat{a},\hat{\varsigma}_{0}^{\sigma,-\sigma}\right]|\psi_{l\sigma}\rangle.
\end{align}
For the term with the commutator we use the explicit form of $\hat{\varsigma}_{0}^{\sigma,-\sigma}$~\eqref{eqs:lefteigenvectors} together with an integral representation for the modified Bessel function $I_n(z)=\int_{-\pi}^\pi\frac{d\phi}{2\pi}\,e^{z\cos\phi+in\phi}$ to find
\begin{subequations}
\label{eqs:projectionOfNon-comp}\begin{align}
  &{}\langle\mathcal{C}_{\alpha}^{-}\vert\left[\hat{a}^{\dagger}\hat{a},\varsigma_{0}^{+-}\right]\vert\mathcal{\psi}_{l+}\rangle\nonumber\\
    {}&{} =\mathcal{N}_0 \sum_{n,m=0}^\infty \frac{(-1)^{n-m}I_{n-m}(\alpha^2)}{(2n)!!\,(2m)!!}\,\frac{\alpha^{2n+1}}{\sqrt{\sinh\alpha^2}}\,\langle0|\hat{a}^{2m}|\psi_{l+}\rangle\nonumber\\
    {}&{} = \mathcal{N}_0\frac{\alpha}{\sqrt{\sinh\alpha^2}}
    \int_{-\pi}^\pi\frac{d\phi}{2\pi}\,
    \langle0|\exp[(\alpha^2-\hat{a}^2)e^{-i\phi}/2]|\psi_{l+}\rangle \nonumber\\
    {}&{} = \mathcal{N}_0\frac{\alpha\,\langle0|\psi_{l+}\rangle}{\sqrt{\sinh\alpha^2}},\\
    {}&{}\langle\mathcal{C}_{\alpha}^+\vert\left[\hat{a}^{\dagger}\hat{a},\varsigma_{0}^{-+}\right]\vert\mathcal{\psi}_{l-}\rangle
    =-\mathcal{N}_0\frac{\langle0|\hat{a}|\psi_{l-}\rangle}{\sqrt{\cosh\alpha^2}},
\end{align}\end{subequations}
where $\mathcal{N}_0=\sqrt{\frac{2\alpha^2}{\sinh2\alpha^2}}$ is a normalization factor.
Upon the substitution of the obtained matrix elements into the expression for $\lambda_z^{(2)}$, it is convenient to split it into two parts as $\lambda_{z}^{(2)}=\kappa_1^2 \left (S_1 + S_2 \right)/\kappa_2$, such that 
\begin{subequations}
\label{eqs:sumsS12viacatstates}
    \begin{align}
    S_{1}=&\sum_{l>0,\sigma}\frac{\langle\psi_{l\sigma}\vert\hat{a}^{\dagger}\hat{a}\vert\mathcal{C}_{\alpha}^{\sigma}\rangle}{4\mu_{l\sigma}}\langle\mathcal{C}_{\alpha}^{-\sigma}\vert\left[\hat{a}^{\dagger}\hat{a},\varsigma_{0}^{\sigma,-\sigma}\right]\vert\psi_{l\sigma}\rangle+\text{h.c.}\nonumber
    \\=&\frac{\alpha^{2}}{\sqrt{2\sinh2\alpha^{2}}}\sum_{l>0}\Biggl(\frac{\alpha\langle0\vert\mathcal{\psi}_{l+}\rangle\langle\mathcal{\psi}_{l+}\vert\hat{a}^{\dagger}\vert\mathcal{C}_{\alpha}^{-}\rangle}{\mu_{l+}\sqrt{\cosh\alpha^{2}}}\nonumber\\
    &-\frac{\langle1\vert\mathcal{\psi}_{l-}\rangle\langle\mathcal{\psi}_{l-}\vert\hat{a}^{\dagger}\vert\mathcal{C}_{\alpha}^{+}\rangle}{\mu_{l-}\sqrt{\sinh\alpha^{2}}}\Biggr),\\
S_{2}=&\sum_{l>0,\sigma}\frac{\langle\psi_{l\sigma}\vert\hat{a}^{\dagger}\hat{a}\vert\mathcal{C}_{\alpha}^{\sigma}\rangle}{2\mu_{l\sigma}}\Bigl(\langle\mathcal{C}_{\alpha}^{-\sigma}|\hat{\varsigma}_{0}^{\sigma,-\sigma}\hat{a}^{\dagger}\hat{a}|\psi_{l\sigma}\rangle\nonumber\\
&-\langle\mathcal{C}_{\alpha}^{-\sigma}\vert\hat{a}^{\dagger}\hat{\varsigma}_{0}^{-\sigma,\sigma}\hat{a}\vert\psi_{l\sigma}\rangle\Bigr)+\text{h.c.}\nonumber\\
=&\frac{\sqrt{2}\alpha^{2}e^{-\alpha^{2}}}{\sqrt{\sinh2\alpha^{2}}}\sum_{l>0}\Biggl(\frac{\left|\langle\mathcal{C}_{\alpha}^{+}\vert\hat{a}\vert\mathcal{\psi}_{l-}\rangle\right|^{2}\sqrt{\coth\alpha^{2}}}{\mu_{l-}}\nonumber
\\&-\frac{\left|\langle\mathcal{C}_{\alpha}^{-}\vert\hat{a}\vert\mathcal{\psi}_{l+}\rangle\right|^{2}\sqrt{\tanh\alpha^{2}}}{\mu_{l+}}\Biggr).
\end{align}
\end{subequations}

To shorten the notation we wrap the summation over Kerr eigenstates into $\hat{H}_{\perp}^{-1}\equiv \sum_{l>0,\sigma}\frac{\vert\mathcal{\psi}_{l\sigma}\rangle\langle\mathcal{\psi}_{l\sigma}\vert}{\mu_{l\sigma}}$ and use the representation of the cat states through the coherent states, then the expressions for $S_1$ and $S_2$ simplify to 
\begin{subequations}\label{eqs:s1s2viacoherentstates}\begin{align}
    &S_1=-\frac{\alpha^{2}e^{\alpha^{2}/2}}{\sinh2\alpha^{2}}\langle0\vert\left(\hat{a}-\alpha\right)\hat{H}_{\perp}^{-1}a^{\dagger}\vert\alpha\rangle, \\
    &S_2=\frac{2\alpha^{2}}{\sinh2\alpha^{2}}\langle-\alpha\vert \hat{a} \hat{H}_{\perp}^{-1}\hat{a}^{\dagger}\vert\alpha\rangle.
\end{align} 
\end{subequations}
It will be seen below, these quantities also appear in the expressions for the bit-flip rate of the frequency detuning and the Z gate. Using the explicit expression for the inverted Kerr Hamiltonian in the basis of coherent states~\eqref{eq:InverseKerr} it is straightforward to evaluate $S_1,S_2$ and we find, \begin{subequations}\label{eqs:sumS1andS2}
\begin{align}
S_{1}&=\frac{\alpha^{2}\left[\text{Shi}(4\alpha^{2})-\text{Shi}(2\alpha^{2})-\coth2\alpha^{2}\text{Chin}(4\alpha^{2})\right]}{\sinh^{2}2\alpha^{2}}\nonumber\\
&\approx-e^{-2\alpha^2}\phantom{a}(\alpha \gg 1),\\
S_{2}&=\frac{\alpha^{2}\left[\left(2\coth2\alpha^{2}-1\right)\text{Chin}(4\alpha^{2})-\text{Shi}(4\alpha^{2})\right]}{\sinh^{2}2\alpha^{2}}\nonumber\\
&\approx -8\alpha^{2}e^{-4\alpha^{2}}\log\alpha\phantom{a}(\alpha \gg 1). 
\end{align}
\end{subequations}
It is tempting to say that the exponentials in these asymptotics directly follow from the matrix elements in the definitions of $S_1$ and $S_2$~\eqref{eqs:s1s2viacoherentstates}, naively assuming the inverse Kerr Hamiltonian to be a local operator. Actually, that is not the case: In Appendix~\ref{app:inverseKerr} we show that the inverse Kerr Hamiltonian is a non-local operator, and thus one can hardly predict the asymptotics of $S_1$ and $S_2$ without the calculation.

\section{Z gate}
\label{app:Zgate}
\subsection{First order}
As before, we write down the action of $\mathcal{L}_1 = -i\epsilon_Z[{\hat{a}^\dagger +\hat{a}}, \hat{\rho} ]$ on the computational subspace and insert resolutions of identity in terms of Kerr Hamiltonian eigenstates, 
\begin{align}
    \label{eq:actionofZgate}&\mathcal{L}_{1}\vert\mathcal{C}_{\alpha}^{\sigma}\rangle\langle\mathcal{C}_{\alpha}^{\sigma'}\vert=-i\epsilon_{Z}\left[\hat{a}+\hat{a}^{\dagger},\vert\mathcal{C}_{\alpha}^{\sigma}\rangle\langle\mathcal{C}_{\alpha}^{\sigma'}\vert\right]=\nonumber \\
    &-i\epsilon_{Z}\Bigl(\alpha\left(\tanh\alpha^{2}\right)^{\sigma/2}\vert\mathcal{C}_{\alpha}^{-\sigma}\rangle\langle\mathcal{C}_{\alpha}^{\sigma'}\vert\nonumber\\
    &+\sum_{l=0}^{\infty}\langle\psi_{l,-\sigma}\vert\hat{a}^{\dagger}\vert\mathcal{C}_{\alpha}^{\sigma}\rangle\vert\psi_{l,-\sigma}\rangle\langle\mathcal{C}_{\alpha}^{\sigma'}\vert\nonumber\\
    &-\alpha\left(\tanh\alpha^{2}\right)^{\sigma'/2}\vert\mathcal{C}_{\alpha}^{\sigma}\rangle\langle\mathcal{C}_{\alpha}^{-\sigma'}\vert\nonumber\\
    &-\sum_{l=0}^{\infty}\langle\mathcal{C}_{\alpha}^{\sigma'}\vert\hat{a}\vert\psi_{l,-\sigma'}\rangle\vert\mathcal{C}_{\alpha}^{\sigma}\rangle\langle\psi_{l,-\sigma'}\vert \Bigr).
\end{align}
To project this action on the cat subspace we pick the $l=0$ terms in the sums and obtain a $4\times4$ matrix. The eigensystem of this matrix has a clear geometrical meaning. Two eigenvalues remain zero, representing conservation of the trace and the $z$ component. The other two eigenvalues are imaginary and complex conjugated: $\lambda_{x\pm iy}^{(1)}=\pm 2i\alpha\epsilon_{Z}\left(\sqrt{\tanh\alpha^{2}}+\sqrt{\coth\alpha^{2}}\right)\approx \pm4i\alpha\epsilon_{Z}\phantom{a}(\alpha\gg1)$, representing rotation around the $z$ axis. Turning on the perturbation for time interval $T$ such that $\epsilon_Z = \pi/(4\alpha T)$, results in the rotation by the angle $\pi$ around $z$ axis, thus implementing the single-qubit Z gate. 
\subsection{Second order}
\begin{figure}
    \centering
    \includegraphics[scale=0.5]{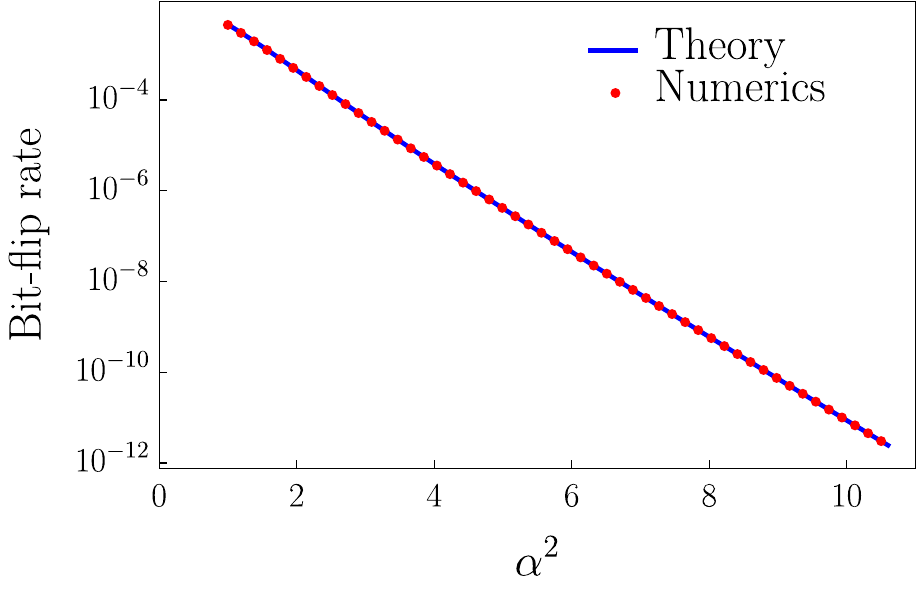}
    \caption{Bit-flip rate in the units of $\kappa_2$ during the Z gate implementation of strength $\epsilon_Z= \kappa_2/10$. The theoretical curve (in blue) represents  $\Gamma_{0\leftrightarrow1}=-\lambda_{z}^{(2)}/2$, where $\lambda_{z}^{(2)}/2$ is given by Eq.~\eqref{eq:zgatefinal}. Note that our formula also shows good agreement with the numerics in the region of intermediate cat-state size $\alpha \gtrsim 1$. }
    \label{fig:bitflipZgate}
\end{figure}

In the first-order calculation we found a two dimensional stable subspace spanned by $\varrho_0^I$ and $\varrho_0^z$ inside the computational subspace, meaning that the degeneracy was not fully resolved. However, acting with $\mathcal{L}_1 = -i\epsilon_Z[{\hat{a}^\dagger +\hat{a}}, \hat{\rho} ]$ twice onto $|\mathcal{C}_\alpha^\sigma\rangle\langle\mathcal{C}_\alpha^{\sigma'}|$ leaves the product of the parities $\sigma\sigma'$ conserved. This guarantees that in the subspace spanned by $\varrho_0^I, \varrho_0^z$ the second-order correction matrix is diagonal. Clearly, the eigenvalue corresponding to $\varrho_0^I$ must be $0$ (trace conservation), while the eigenvalue corresponding to $\varrho_0^z$ is the exponentially small bit-flip rate. Using the action of $\mathcal{L}_1$  \eqref{eq:actionofZgate} and proceeding like with the single photon loss, we find for the exponentially small eigenvalue, 
\begin{align}
  \lambda_z^{(2)}&=-\text{Tr}\Biggl\{\frac{\varsigma_{0}^{\sigma,-\sigma}+\varsigma_{0}^{-\sigma,\sigma}}{2}\mathcal{L}_{1}\mathcal{L}_{0}^{-1}\mathcal{P}_{\perp}\mathcal{L}_{1}\sum_{\sigma}\vert\mathcal{C}_{\alpha}^{\sigma}\rangle\langle\mathcal{C}_{\alpha}^{-\sigma}\vert\Biggr\}\nonumber\\
  &=-\epsilon_{Z}^{2}\sum_{\sigma,l>0}\frac{\langle\psi_{l\sigma}\vert\hat{a}^{\dagger}\vert\mathcal{C}_{\alpha}^{-\sigma}\rangle}{\kappa_2\mu_{l\sigma}}\nonumber\\
  &\times\langle\mathcal{C}_{\alpha}^{\sigma}\vert\left[{\varsigma_{0}^{\sigma,-\sigma}+\varsigma_{0}^{-\sigma,\sigma}},\hat{a}+\hat{a}^{\dagger}\right]\vert\psi_{l\sigma}\rangle+\text{h.c.}.
\end{align}
Instead of directly calculating the matrix elements with the left eigenvector, we highlight the correspondence to the expression \eqref{eq:secondorderforloss_intermediateApp} for the second-order contribution due to the single photon loss. To do that we firstly simplify the last matrix element, 
\begin{align}
    &\langle\mathcal{C}_{\alpha}^{\sigma}\vert\left[{\varsigma_{0}^{\sigma,-\sigma}+\varsigma_{0}^{-\sigma,\sigma}},\hat{a}+\hat{a}^{\dagger}\right]\vert\psi_{l\sigma}\rangle \nonumber\\
    &=\langle\mathcal{C}_{\alpha}^{\sigma}\vert\varsigma_{0}^{-\sigma,\sigma}\left(\hat{a}+\hat{a}^{\dagger}\right)\vert\psi_{l\sigma}\rangle-\langle\mathcal{C}_{\alpha}^{\sigma}\vert\left(\hat{a}+\hat{a}^{\dagger}\right)\varsigma_{0}^{\sigma,-\sigma}\vert\psi_{l\sigma}\rangle\nonumber\\
    &=\langle\mathcal{C}_{\alpha}^{\sigma}\vert\varsigma_{0}^{-\sigma,\sigma}\hat{a}\vert\psi_{l\sigma}\rangle-\langle\mathcal{C}_{\alpha}^{\sigma}\vert\hat{a}\varsigma_{0}^{\sigma,-\sigma}\vert\psi_{l\sigma}\rangle.
\end{align}
In both terms $\hat{a}^\dagger$ drops out. Indeed, in the second term it produces scalar product with the right eigenvector  $\langle\mathcal{C}_{\alpha}^{-\sigma}\vert\varsigma_{0}^{\sigma,-\sigma}\vert\psi_{l\sigma}\rangle=0$. In the first term we proceed as,
\begin{align}
    &\langle\mathcal{C}_{\alpha}^{\sigma}\vert\varsigma_{0}^{-\sigma,\sigma}\left(\hat{a}+\hat{a}^{\dagger}\right)\vert\psi_{l\sigma}\rangle\nonumber\\
    &=\langle\mathcal{C}_{\alpha}^{\sigma}\vert\varsigma_{0}^{-\sigma,\sigma}\sum_{l',\sigma'}\vert\psi_{l'\sigma'}\rangle\langle\psi_{l'\sigma'}\vert\left(\hat{a}+\hat{a}^{\dagger}\right)\vert\psi_{l\sigma}\rangle\nonumber\\
    &=\langle\mathcal{C}_{\alpha}^{\sigma}\vert\varsigma_{0}^{-\sigma,\sigma}\vert\mathcal{C}_{\alpha}^{-\sigma}\rangle\langle\mathcal{C}_{\alpha}^{-\sigma}\vert\left(\hat{a}+\hat{a}^{\dagger}\right)\vert\psi_{l\sigma}\rangle\nonumber\\
    &=\langle\mathcal{C}_{\alpha}^{\sigma}\vert\varsigma_{0}^{-\sigma,\sigma}\vert\mathcal{C}_{\alpha}^{-\sigma}\rangle\langle\mathcal{C}_{\alpha}^{-\sigma}\vert\hat{a}\vert\psi_{l\sigma}\rangle=\langle\mathcal{C}_{\alpha}^{\sigma}\vert\varsigma_{0}^{-\sigma,\sigma}\hat{a}\vert\psi_{l\sigma}\rangle.
\end{align}
In order to make the expression look like Eq. \eqref{eq:secondorderforloss_intermediateApp} we also insert $\hat{a}^\dagger$ to act on the cat states at the cost of a common factor $1/\alpha^2$ in the eigenvalue contribution, 
\begin{align}
    &\lambda_z^{(2)}=-\frac{\epsilon_{Z}^{2}}{\kappa_2\alpha^{2}}\sum_{\sigma,l>0}\frac{\langle\psi_{l\sigma}\vert\hat{a}^{\dagger}\hat{a}\vert\mathcal{C}_{\alpha}^{-\sigma}\rangle}{\mu_{l\sigma}}\times\nonumber\\
    &(\langle\mathcal{C}_{\alpha}^{-\sigma}\vert\hat{a}^{\dagger}\varsigma_{0}^{-\sigma,\sigma}\hat{a}\vert\psi_{l\sigma}\rangle-\langle\mathcal{C}_{\alpha}^{-\sigma}\vert\hat{a}^{\dagger}\hat{a}\varsigma_{0}^{\sigma,-\sigma}\vert\psi_{l\sigma}\rangle) +\text{h.c.}.
    \label{eq:ZgateIntermediat}
\end{align}
Actually, this expression is a linear combination of the introduced sums $S_1$ and $S_2$ \eqref{eqs:sumsS12viacatstates}. To show this we represent the last term as a sum of an anticommutator and a commutator, 
\begin{align}
    &\langle\mathcal{C}_{\alpha}^{-\sigma}\vert\hat{a}^{\dagger}\hat{a}\varsigma_{0}^{\sigma,-\sigma}\vert\psi_{l\sigma}\rangle = {}{} \frac{1}{2}\langle\mathcal{C}_{\alpha}^{-\sigma}\vert\left\{ \hat{a}^{\dagger}\hat{a},\varsigma_{0}^{\sigma,-\sigma}\right\} \vert\psi_{l\sigma}\rangle\nonumber
    \\
    &+\frac{1}{2}\langle\mathcal{C}_{\alpha}^{-\sigma}\vert\left[\hat{a}^{\dagger}\hat{a},\varsigma_{0}^{\sigma,-\sigma}\right]\vert\psi_{l\sigma}\rangle.
\end{align}
The anticommutator term together with the first term in Eq.~\eqref{eq:ZgateIntermediat} reduces to the expression for the second-order contribution due to the single photon loss \eqref{eq:secondorderforloss_intermediateApp}. The commutator term produces the output proportional to $S_1$ as was shown before. All in all, we find for the bit-flip rate of the Z gate,
\begin{align}\label{eq:zgatefinal}
\lambda_z^{(2)}&=\frac{2\epsilon_{Z}^{2}}{\kappa_2\alpha^{2}}(2S_1+S_2) \nonumber\\
&=\frac{2\epsilon_{Z}^{2}\left[\text{Shi}(4\alpha^{2})-\text{Chin}(4\alpha^{2})-2\text{ Shi}(2\alpha^{2})\right]}{\kappa_{2}\sinh^{2}2\alpha^{2}}\nonumber\\
&\approx -\frac{4\epsilon_{Z}^{2}}{\kappa_2\alpha^{2}}e^{-2\alpha^{2}}\phantom{a}(\alpha \gg 1).
\end{align}
To support our analytical finding, we compare this bit-flip rate to the result of the exact diagonalization of the perturbed Lindbladian (see Fig.~\ref{fig:bitflipZgate}).

\section{Frequency detuning
}
\label{app:detuning}
\subsection{First order}
In this Appendix we derive first- and second-order contributions to the computational subspace eigenvalues due to the detuning perturbation $\mathcal{L}_1=-i\Delta[\hat{a}^\dagger \hat{a}, \hat{\rho}]$. To begin with, we write down the action of the detuning on the computational subspace
\begin{align}
\mathcal{L}_{1}\vert\mathcal{C}_{\alpha}^{\sigma}\rangle\langle\mathcal{C}_{\alpha}^{\sigma^{\prime}}\vert={}&{} -i\Delta\left[\hat{a}^{\dagger}\hat{a},\vert\mathcal{C}_{\alpha}^{\sigma}\rangle\langle\mathcal{C}_{\alpha}^{\sigma^{\prime}}\vert\right]\nonumber\\
={}&{} -i\Delta\sum_{l=0}^{\infty}\langle\mathcal{\psi}_{l\sigma}\vert\hat{a}^{\dagger}\hat{a}\vert\mathcal{C}_{\alpha}^{\sigma}\rangle\vert\mathcal{\psi}_{l\sigma}\rangle\langle\mathcal{C}_{\alpha}^{\sigma^{\prime}}\vert\nonumber\\
{}&{} +i\Delta\sum_{l=0}^{\infty}\langle\mathcal{C}_{\alpha}^{\sigma^{\prime}}\vert\hat{a}^{\dagger}\hat{a}\vert\mathcal{\psi}_{l\sigma^{\prime}}\rangle\vert\mathcal{C}_{\alpha}^{\sigma}\rangle\langle\mathcal{\psi}_{l\sigma^{\prime}}\vert.
\label{eq:detuning_action}
\end{align}
Again we inserted resolutions of identity in terms of the Kerr eigenstates $\hat{I}=\sum_{l\sigma}\vert\psi_{l\sigma}\rangle\langle\psi_{l\sigma}\vert$ to immediately obtain an expansion in the right eigenvectors of $\mathcal{L}_0$. The projection onto the computational subspace corresponds to picking $l=0$ terms in the sums and we get
\begin{equation}
\mathcal{P}_\|\mathcal{L}_1|\mathcal{C}_\alpha^\sigma\rangle\langle\mathcal{C}_\alpha^{\sigma'}|=i\Delta\frac{\alpha^2}{\sinh2\alpha^2}(\sigma-\sigma')|\mathcal{C}_\alpha^\sigma\rangle\langle\mathcal{C}_\alpha^{\sigma'}|.
\end{equation}
The resulting $4\times 4$ matrix is diagonal, which is an obvious consequence of the parity conservation. Two eigenvalues $\lambda_{+-}=\lambda^*_{-+}$ become finite, corresponding to the components $|\mathcal{C}_\alpha^+\rangle\langle\mathcal{C}_\alpha^{-}|$ and $|\mathcal{C}_\alpha^-\rangle\langle\mathcal{C}_\alpha^{+}|$ inside the cat subspace. Where as the two other eigenvalues, corresponding to  $|\mathcal{C}_\alpha^+\rangle\langle\mathcal{C}_\alpha^{+}|$ and $|\mathcal{C}_\alpha^-\rangle\langle\mathcal{C}_\alpha^{-}|$,  remain identically $0$ in each perturbation order. Indeed, note that the zeroth-order left eigenvectors $\varsigma^{++}_0$ and $\varsigma^{--}_0$ are exact left eigenvectors of the Lindbladian $\mathcal{L}_0 + \mathcal{L}_1$. To conclude, the first-order contributions due to the detuning are 
\begin{equation}
\lambda_{+-}^{(1)}=-\lambda_{-+}^{(1)}=i\Delta\frac{2\alpha^2}{\sinh2\alpha^2}.
\end{equation}
To sum up, the detuning perturbation in the first order induces exponentially slow rotation over the $x$ axis of the Bloch sphere.

\subsection{Second order}
\begin{figure}
    \centering
    \includegraphics[scale=0.5]{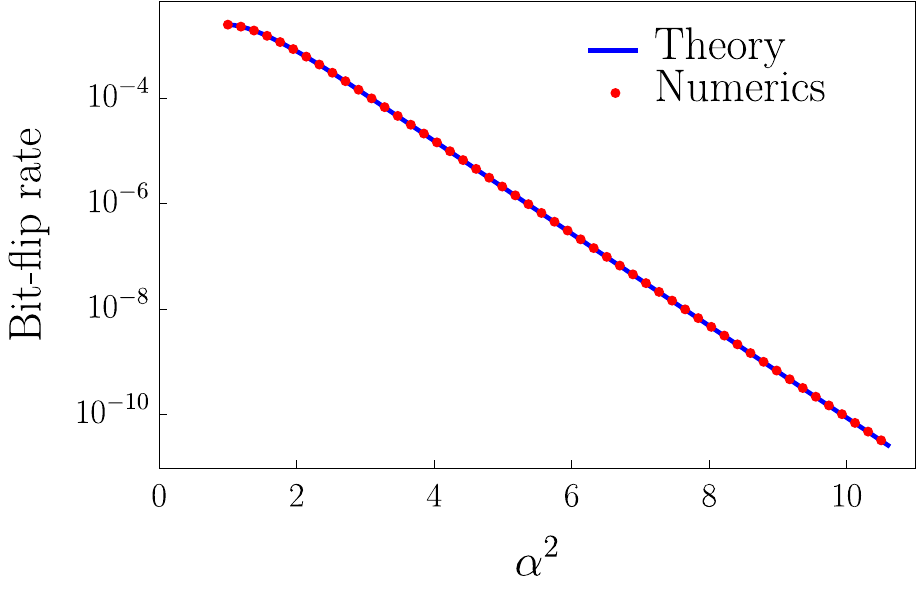}
    \caption{Bit-flip rate in the units of $\kappa_2$ due to the frequency detuning of strength $\Delta = \kappa_2/10$. The theoretical curve (in blue) represents  $\Gamma_{0\leftrightarrow1}=-\lambda_{+-}^{(2)}/2$, where $\lambda_{+-}^{(2)}/2$ is given by Eq.~\eqref{eq:detuningfinalresult}. Note that our formula also shows good agreement with the numerics in the region of intermediate cat-state size $\alpha \gtrsim 1$. }
    \label{fig:bitflipDetuning}
\end{figure}

For the second-order correction we need to evaluate $\lambda^{(2)}_{+-}=-\Tr \left(\varsigma_0^{+-}\mathcal{L}_1\mathcal{L}_0^{-1}\mathcal{P}_\perp\mathcal{L}_1|\mathcal{C}_\alpha^+\rangle\langle\mathcal{C}_\alpha^{-}|\right) $. Projection $\mathcal{P}_\perp\mathcal{L}_1\hat\rho_\|$ corresponds to picking the $l>0$ terms in Eq.~\eqref{eq:detuning_action}. The action of $\mathcal{L}_0^{-1}$ reduces to simply dividing by the eigenvalues given in Eq.~(\ref{eqs:Lindbladian_eigenvectorsl00l}). Then we have to apply $\mathcal{L}_1$ for the second time and project the result on the computational subspace, which was already done in Eqs. \eqref{eqs:projectionOfNon-comp}.
This leads to the following expression for the second-order correction
\begin{align}
   \lambda_{+-}^{(2)} &=\frac{\Delta^{2}}{\kappa_2}\frac{\sqrt{8}\alpha^{2}}{\sqrt{\sinh2\alpha^{2}}}\sum_{l>0}\biggl(\frac{\alpha\langle0|\psi_{l+}\rangle\langle\psi_{l+}|\hat{a}^{\dagger}|\mathcal{C}_{\alpha}^{-}\rangle}{\mu_{l+}\sqrt{\cosh\alpha^{2}}}\nonumber\\
   &-\frac{\langle1|\psi_{l-}\rangle\langle\psi_{l-}|\hat{a}^{\dagger}|\mathcal{C}_{\alpha}^{+}\rangle}{\mu_{l-}\sqrt{\sinh\alpha^{2}}}\biggr)
   \label{eq:ReLambda_eigensystem}
\end{align}
in terms of the Kerr Hamiltonian eigensystem. By comparison with Eqs. \eqref{eqs:sumsS12viacatstates}, we immediately notice that
\begin{align} \label{eq:detuningfinalresult}
    &\lambda_{+-}^{(2)} = \lambda_{-+}^{(2)} = 4 \Delta^2 S_1/\kappa_2 \nonumber\\ 
    &=\frac{4\Delta^2\alpha^{2}\left[\text{Shi}(4\alpha^{2})-\text{Shi}(2\alpha^{2})-\coth2\alpha^{2}\text{Chin}(4\alpha^{2})\right]}{\kappa_2\sinh^{2}2\alpha^{2}}\nonumber\\
    &\approx  -\frac{4\Delta^2}{\kappa_2}e^{-2\alpha^{2}} \phantom{a} (\alpha \gg 1).
\end{align} (since the second-order contribution is real).  In terms of the Bloch sphere this result translates into simultaneous decay of $y$ and $z$ components of the qubit density matrix with the rate $\lambda_{+-}^{(2)} = \lambda_{-+}^{(2)}$. This result agrees with the real-time instanton calculation of Ref.~\cite{Thompson2022} at $\alpha \gg 1$. Moreover, the present calculation yields an analytical expression for the pre-exponential factor. To further support our analytical finding we compare our bit-flip rate to the result of the exact diagonalization of the perturbed Lindbladian (see Fig.~\ref{fig:bitflipDetuning}).

\section{Inverse Kerr Hamiltonian}
\label{app:inverseKerr}
In this Appendix we present the explicit form of the matrix element of the inverse Kerr Hamiltonian in the basis of unnormalized coherent states. To find the solution we assume that the variables $\chi$ and $\varphi$ are real and lie in the domain $-\alpha\leq \chi,\varphi\leq \alpha$. Then our initial differential equation~\eqref{eq:diffeq} transforms into a non-homogeneous boundary value problem~\cite{Polyanin+Zaitsev}. The Green's function for this problem is
\begin{align}
    g(\chi, t) = \frac{\cosh (\alpha  (t+\chi))-\cosh (\alpha  (|t-\chi| -2 \alpha ))}{2 \alpha  \sinh 2 \alpha ^2}.  
\end{align}
Thus we can write the solution in the limited domain as
\begin{align}
    &H_{\perp}^{-1}(\chi,\varphi)=\intop_{-\alpha}^{\alpha}\frac{g(\chi, t)dt}{t^2-\alpha^2}\langle\chi\vert(1-\sum_{\sigma}\vert\mathcal{C}_{\alpha}^{\sigma}\rangle\langle\mathcal{C}_{\alpha}^{\sigma}\vert)\vert\varphi\rangle.  
\end{align}
The result of the integration can be represented via the complementary exponential integral $\text{Ein}(z)$~\cite{abramowitz+stegun}, which is an holomorphic function in the complex plane of $z$. Therefore, the analytical continuation for arbitrary complex $\chi$ and $\varphi$ is straightforward and the final result for the inverse Kerr Hamiltonian is, 
\begin{widetext}
\begin{align} \label{eq:InverseKerr}
H_{\perp}^{-1}(\bar{\chi},\varphi)
&=\frac{1}{16\alpha^{2}}\Biggl\{\frac{1}{\sinh2\alpha^{2}}\Bigl[\text{Ein}(-4\alpha^{2})\left(4e^{-3\alpha^{2}}\cosh(\alpha\varphi-\alpha\bar{\chi})-2e^{-5\alpha^{2}}\cosh(\alpha\bar{\chi}+\alpha\varphi)-2e^{-\alpha^{2}}\cosh(\alpha\bar{\chi}+\alpha\varphi)\right)\nonumber\\
&+\nonumber\text{Ein}(4\alpha^{2})\left(2e^{\alpha^{2}}\cosh(\alpha\varphi-\alpha\bar{\chi})-4e^{3\alpha^{2}}\cosh(\alpha\bar{\chi}+\alpha\varphi)+2e^{5\alpha^{2}}\cosh(\alpha\varphi-\alpha\bar{\chi})\right)\Bigr]\nonumber\\
&+\frac{4}{\sinh2\alpha^{2}}\Bigl[e^{-\alpha\bar{\chi}}\sinh(\alpha^{2}+\alpha\varphi)\left(e^{-2\alpha^{2}}\text{Ein}(-2\alpha^{2}-2\alpha\bar{\chi})-e^{2\alpha^{2}}\text{Ein}(2\alpha^{2}-2\alpha\bar{\chi})\right)\nonumber\\
&+e^{\alpha\bar{\chi}}\sinh(\alpha^{2}-\alpha\varphi)\left(e^{-2\alpha^{2}}\text{Ein}(2\alpha\bar{\chi}-2\alpha^{2})-e^{2\alpha^{2}}\text{Ein}(2\alpha^{2}+2\alpha\bar{\chi})\right)\nonumber\\
&+e^{-\alpha\varphi}\sinh(\alpha^{2}+\alpha\bar{\chi})\left(e^{-2\alpha^{2}}\text{Ein}(-2\alpha^{2}-2\alpha\varphi)-e^{2\alpha^{2}}\text{Ein}(2\alpha^{2}-2\alpha\varphi)\right)\nonumber 
\\ 
&+e^{\alpha\varphi}\sinh(\alpha^{2}-\alpha\bar{\chi})\left(e^{-2\alpha^{2}}\text{Ein}(2\alpha\varphi-2\alpha^{2})-e^{2\alpha^{2}}\text{Ein}(2\alpha^{2}+2\alpha\varphi)\right)\Bigr]\nonumber\\
&+4e^{\alpha^{2}}\left(e^{\alpha\bar{\chi}-\alpha\varphi}\text{Ein}(\alpha^{2}+\alpha\bar{\chi}-\alpha\varphi-\bar{\chi}\varphi)+e^{\alpha\varphi-\alpha\bar{\chi}}\text{Ein}(\alpha^{2}-\alpha\bar{\chi}+\alpha\varphi-\bar{\chi}\varphi)\right)\nonumber\\
&-4e^{-\alpha^{2}}\left(e^{\alpha\bar{\chi}+\alpha\varphi}\text{Ein}(-\alpha^{2}+\alpha\bar{\chi}+\alpha\varphi-\bar{\chi}\varphi)+e^{-\alpha\bar{\chi}-\alpha\varphi}\text{Ein}(-\alpha^{2}-\alpha\bar{\chi}-\alpha\varphi-\bar{\chi}\varphi)\right) \Biggr\}.
\end{align}
\end{widetext}

\bibliography{references}

\end{document}